\documentclass[12pt]{article}
\pdfoutput=1 % if your are submitting a pdflatex (i.e. if you have images in pdf, png or jpg format)
\usepackage{amsmath}
\usepackage{amssymb}
\usepackage{graphicx}
\usepackage{enumerate}
\usepackage{slashed, bm}
\usepackage{hyperref, xcolor}
\definecolor{dark}{rgb}{0.10,0.2,0.3}
\definecolor{magenta}{rgb}{0.7,0.1,0.3}
\definecolor{purpure}{rgb}{0.5,0.15,0.3}
\usepackage[font=small,format=plain,labelfont=bf,up,textfont=it,up]{caption}
\usepackage{hyperref, cite}
\hypersetup{colorlinks,
citecolor=blue,
filecolor=blue,
linkcolor=magenta,
urlcolor=purpure,hyperfootnotes=true,pdftex}

\usepackage[normalem]{ulem}
\usepackage{subfig}
\newcommand{\beq}{\begin{equation}}
\newcommand{\eeq}{\end{equation}}
\newcommand{\bea}{\begin{eqnarray}}
\newcommand{\eea}{\end{eqnarray}}
\def\cascade{{\sc Cascade}}
\def\katie{\mbox{\sc Ka\hspace{-0.2ex}Tie}}
\def\pythia{{\sc Pythia}}

\title{%
\vspace{-4ex}{\normalsize\normalfont\hspace{\fill}DESY 18-159}\\
\vspace{-1ex}{\normalsize\normalfont\hspace{\fill}IFJPAN-IV-2018-16}\\
\vspace{2ex}%
\boldmath \large \bf Calculation of the $Z+\mathrm{jet}$ cross section including transverse momenta of initial partons}

\author{M.~Deak${}^a$, A.~van~Hameren${}^a$, H.~Jung${}^b$,\\
        A.~Kusina${}^{a}$, K.~Kutak${}^{a,d}$, M.~Serino${}^{c}$
\bigskip \\
${}^a$ Institute of Nuclear Physics, Polish Academy of Sciences, \\ ul.~Radzikowskiego 152, 31-342, Cracow,
Poland\\
${}^b$ DESY, Hamburg, Germany\\
${}^c$ Department of Physics, \\  Ben Gurion University of the
Negev, \\ Beer Sheva 8410501, Israel\\
${}^d$ Theoretical Physics Department, CERN,\\
1211 Geneva 23, Switzerland
}

\begin{document}

\maketitle \flushbottom
\begin{abstract}
{We perform calculations of $Z$+jet cross-section taking into account the transverse momenta of the initial partons. Transverse Momentum Dependent (TMD) parton densities obtained with the Parton Branching method are used and higher order corrections are included via TMD parton showers in the initial state.
The predictions are compared to measurements of forward $Z$+jet production of the LHCb collaboration at $\sqrt{s}=7$~TeV.
We show that the results obtained in $k_T$-factorization are in good agreement with results obtained from a NLO calculation matched with traditional parton showers. We also demonstrate that in the forward rapidity region, $k_T$-factorization and hybrid factorization predictions agree with each other.
}
\end{abstract}

%\tableofcontents
\newpage
{
  \hypersetup{linkcolor=black}
%  \tableofcontents
}

%\newpage

%%%%%%%%%%%%%%%%%%%%%%%%%%%%%%%%%%%%%%%%%%%%%%%%%%%%%%
\section{Introduction}
\label{sec:intro}
%%%%%%%%%%%%%%%%%%%%%%%%%%%%%%%%%%%%%%%%%%%%%%%%%%%%%%

The Large Hadron Collider (LHC) opened opportunities to explore kinematic regions where particles
produced in high-energy collisions possess large transverse momenta and a wide range of available rapidities. 
The production of electroweak bosons and jets is a vital test of the Standard Model. 
Furthermore, studies of the associated production of electroweak bosons and jets provide important insights into the transverse
partonic structure of hadrons. In particular, motivated by earlier studies~\cite{Hautmann:2012sh,Motyka:2014lya,vanHameren:2015uia,Schafer:2016nul,Celiberto:2018muu}, one can use recent 
theoretical and technical advancements to study in more detail Transverse Momentum Dependent Parton Densities (TMDs)~\cite{Angeles-Martinez:2015sea}. 

Final states of the $Z$+jet type, being a combination of colored and colourless partons and particles, give the opportunity for
investigations which complement the results obtained in studies of pure jet final states~\cite{Deak:2010gk}. 
This is because final state rescatterings due to soft color exchanges 
have less impact on the properties of the produced final state as compared to pure jet final states. 

This work focuses on predictions using $k_T$-factorization \cite{Catani:1990eg} (also referred to as High Energy Factorization) as implemented in the parton-level event generator \katie \cite{vanHameren:2016kkz} combined with a TMD initial state parton shower implementation in new version of \cascade\ ~\cite{Jung:2010si,Bury:2017jxo} compared to $Z$+jet measurements of the LHCb collaboration~\cite{Aaij:2013nxa}.
%in the forward direction within the muon rapidity range $2.0<\eta^\mu<4.5$.
%\comment{How do we account for this finite muon rapidity range in our calculation, which stops at the Z boson ? }
%The final-state muon and anti-muon were required to have transverse momenta $p_T^{\mu}>20$ GeV,
%the mass of the produced muon pair between $60<M_{\mu\mu}<120$ GeV
%while the leading jet is required to have pseudorapidity of $2.0<\eta^{\mathrm{jet}}<4.5$, and the leading jet
%must be separated from the muons by $\Delta r(\mu,\mathrm{jet})>0.4$. Additionally the cross-sections were
%measured for two thresholds of the leading jet transverse momentum of $20$ and $10$ GeV.

The $k_T$-factorization formula for the inclusive cross section schematically reads
\beq
\label{eq:fac}
\sigma=\int \frac{1}{F}dPS\sum_{i,j}{\cal A}_i(x_1, k_{T1}, \mu_F )\otimes|ME|^2\otimes{\cal A}_j(x_2, k_{T2}, \mu_F ) \, ,
\eeq
where $F$ is the flux, $PS$ is the final state phase space and $ME$ is the partonic matrix element. The TMD parton distributions 
${\cal A}_i(x, k_T, \mu_F)$ depend
for a  parton of type $i$ on the longitudinal momentum fraction $x$,  the factorization scale $\mu_F$ and the transverse momentum $k_T$, i.e.\ the momentum perpendicular to the collision axis of the colliding partons.   The formula is valid when  $x$ is not too small and additional effects from gluon recombinations can be neglected~\cite{Kotko:2015ura,vanHameren:2016ftb}.
The matrix elements in the formula above are efficiently calculated numerically using helicity methods \cite{vanHameren:2012uj,vanHameren:2012if} and recursion relation 
as implemented in \katie~\cite{vanHameren:2016kkz}, giving the same results as 
 using Lipatov's effective action~\cite{Antonov:2004hh}.

The TMD parton densities can be defined by introducing
operators whose expectation values count the number of partons and their evolution is given by 
renormalization of divergences either in rapidity or transversal momentum.
One can also construct TMDs starting from the collinear parton densities by ``unfolding'' them 
by the Watt-Kimber-Martin-Ryskin prescription~\cite{Kimber:2001sc,Martin:2009ii}. In the Parton Branching (PB) method~\cite{Hautmann:2017fcj,Hautmann:2017xtx}, which will be used in this article, the TMD is constructed by a parton branching algorithm to solve the DGALP evolution equations. The starting distributions are
obtained from a fit to inclusive DIS cross section measurements \cite{Martinez:2018jxt}, using NLO splitting functions.
The advantage of using PB-TMDs is that once integrated over the transverse momentum the collinear parton density functions are obtained, which is essential for a consistent comparison between $k_T$-factorization and collinear approaches.
The PB-TMDs are well suited to construct initial state parton showers \cite{Bury:2017jxo} in such a way, that the kinematics of the hard process are fixed and no kinematic corrections are applied when the parton shower is added,
as opposed to what is typically done in genuine collinear physics based Monte Carlo generators.
With the PB-TMDs one can address observables initiated by initial state off-shell quarks and,
in the end, perform a Monte Carlo simulation of the whole event, e.g.\ for a process
\beq
%P_A+P_B\longrightarrow Z+j + X \rightarrow \mu^{+} \mu^{-}+j  +  X\, ,
P_A+P_B\longrightarrow Z/\gamma^*+j + X \rightarrow \mu^{+} \mu^{-}+j  +  X \, ,
\eeq
where $P_A$ and $P_B$ are the colliding protons which produce the intermediate state 
with one jet $j$ and one $Z/\gamma^*$ boson and other unobserved particles in a state $X$. 
We compare the prediction with a measurement obtained by LHCb~\cite{Aaij:2013nxa}, where the $Z$-boson decay into muon pairs was studied.

%%%%%%%%%%%%%%%%%%%%%%%%%%%%%%%%%%%%
\section{Results}
\label{sec:results}
%%%%%%%%%%%%%%%%%%%%%%%%%%%%%%%%%%%%

We present results computed in the framework of $k_T$-factorization and Hybrid
Factorization and compare them to the recent $Z+\mathrm{jet}$ measurements of LHCb~\cite{Aaij:2013nxa}.
We use \katie~\cite{vanHameren:2016kkz} in order to produce parton-level events with off-shell
initial-state momenta. These events are then further processed by CASCADE~\cite{Jung:2010si} with its extension for a complete initial state parton shower for all flavours, as described in Ref.~\cite{Bury:2017jxo}. The final state parton shower and 
hadronization are performed by PYTHIA~\cite{Sjostrand:2006za}. 
%The initial state shower of CASCADE is suited for the needs of the $k_T$-factorization framework,  taking into account the transverse momentum of the initial partons given by the used unintegrated PDFs.
% The initial state shower in CASCADE uses the transverse momentum available from the PB PDFs,
% whereas final state shower and hadronization are delegated to PYTHIA~\cite{Sjostrand:2000wi}.
%In this study we employ the parton branching (PB) TMDs~\cite{Martinez:2018jxt}
%(more specifically PB-NLO-2018-Set 2), which are available via the TMDlib library~\cite{Hautmann:2014kza}.
%For all calculations we use two-loop $\alpha_s$ with $\alpha_s(m_Z) = 0.118$.
% All the results that are presented here use parton branching (PB) PDFs
% set 2~\cite{Martinez:2018jxt} which provide access to transverse momentum of partons.
%For results obtained within the hybrid approach (with one on-shell and one
%off-shell initial parton) or within the collinear approach we use the HERAPDF20\_NLO\_EIG collinear
%PDFs~\cite{Abramowicz:2015mha} which are the same as the $k_T$-integrated PB-NLO-2018, by construction.
We also use calculations performed in collinear factorization, with parton showers added, in LO for $Z$+1 jet and at NLO for $Z$+2 jets using POWHEG~\cite{Jager:2012uq} together with \pythia\ \cite{Sjostrand:2014zea} for parton showering, multi-parton interaction and hadronization.
For all calculations we use two-loop $\alpha_s$ with $\alpha_s(m_Z) = 0.118$. The parton branching TMDs~\cite{Martinez:2018jxt}
(more specifically PB-NLO-2018-Set 2), available in TMDlib~\cite{Hautmann:2014kza}, are used for $k_T$ dependent calculations while for the hybrid approach (with one on-shell and one off-shell initial parton) and the collinear approach we use the HERAPDF20\_NLO\_EIG collinear
PDFs~\cite{Abramowicz:2015mha}. The PB-NLO-2018 TMDs, once integrated over $k_T$, reproduce HERAPDF20\_NLO, by construction.
The analysis of the final state particles  is performed with the  Rivet framework~\cite{Buckley:2010ar}.

From all the distributions measured by LHCb~\cite{Aaij:2013nxa} we will concentrate on the
$\Delta\phi$ and $p_T^Z$ distributions, for which going beyond the collinear approximation is most relevant.

%%%%%%%%%%%%%%%%%
\subsection{Parton level results}
%%%%%%%%%%%%%%%%%

We start by discussing results of calculations performed at the parton level (without any showering
or hadronization, but with the decay $Z/\gamma^* \to \mu^+ \mu^-$), by convolving the TMD densities with the matrix elements, as schematically described by Eq.~\eqref{eq:fac}.

%----------------
\begin{figure*}[!tb]
\centering{}
\subfloat[$\Delta\phi$ \normalfont{for} $p_T^{\mathrm{jet}}>10$ \normalfont{GeV}]{
\includegraphics[width=0.48\textwidth]{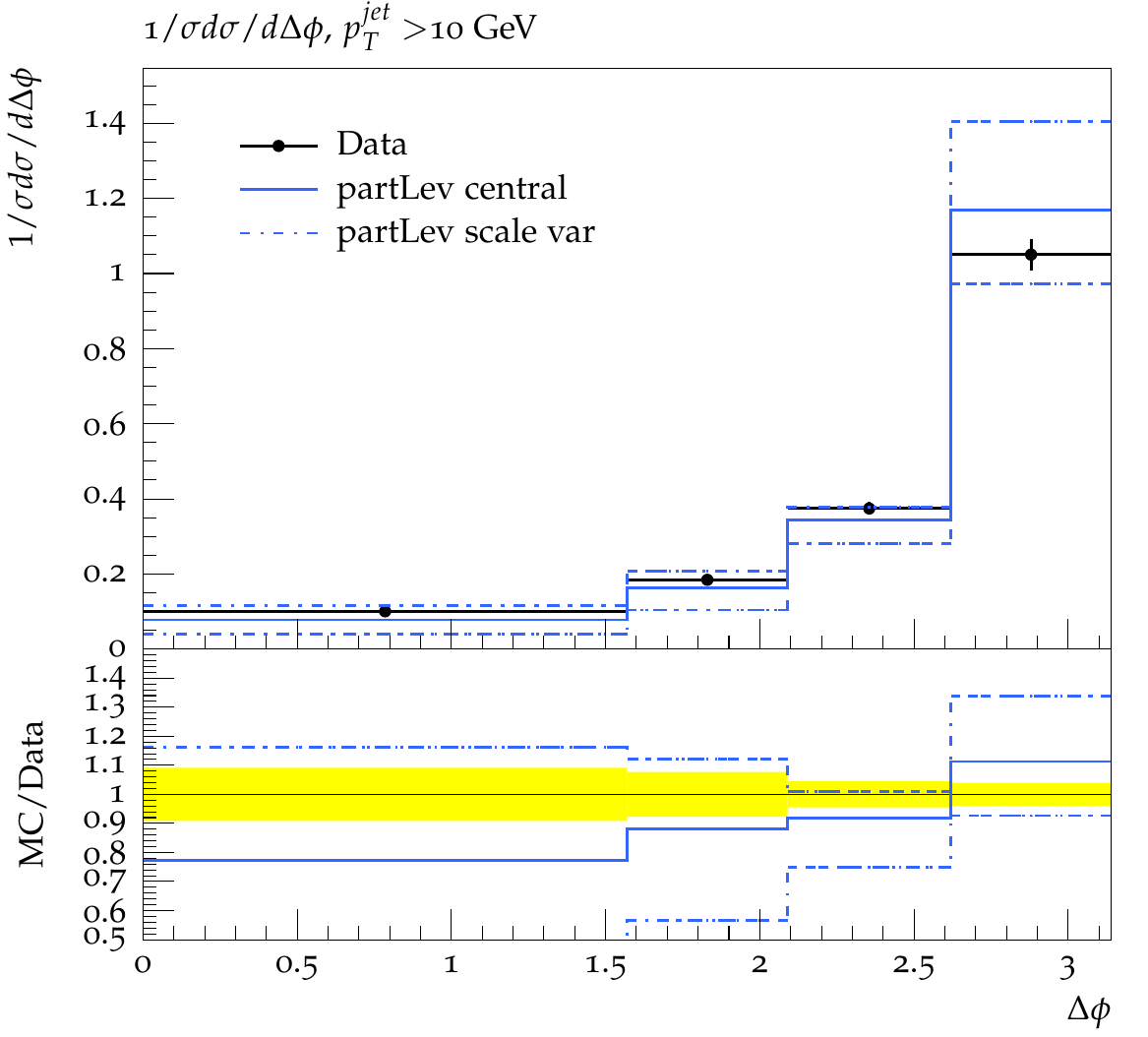}}
\hfil
\subfloat[$\Delta\phi$ \normalfont{for} $p_T^{\mathrm{jet}}>20$ \normalfont{GeV}]{
\includegraphics[width=0.48\textwidth]{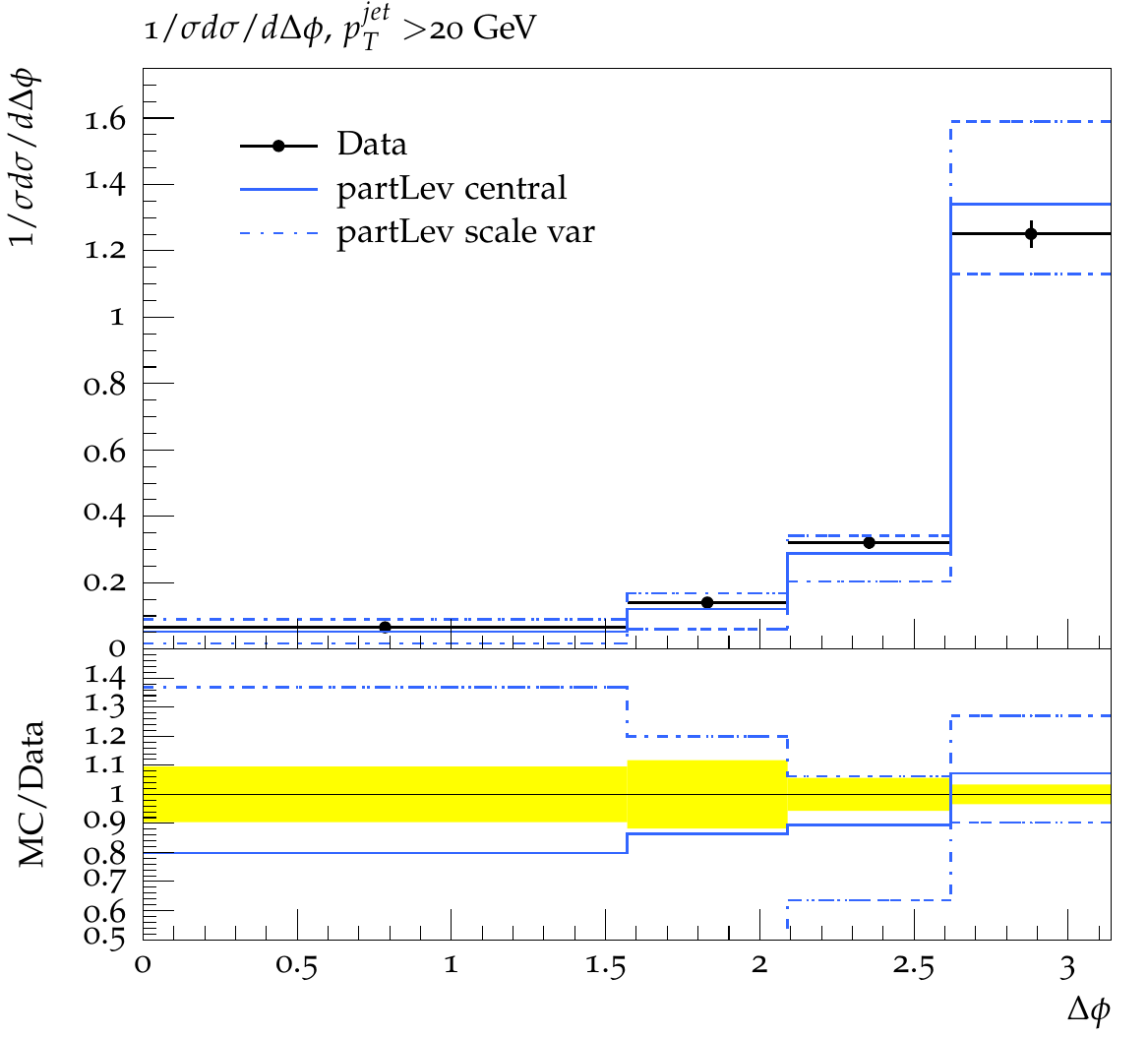}}
\\
\subfloat[$p_T^Z$ \normalfont{for} $p_T^{\mathrm{jet}}>10$ \normalfont{GeV}]{
\includegraphics[width=0.48\textwidth]{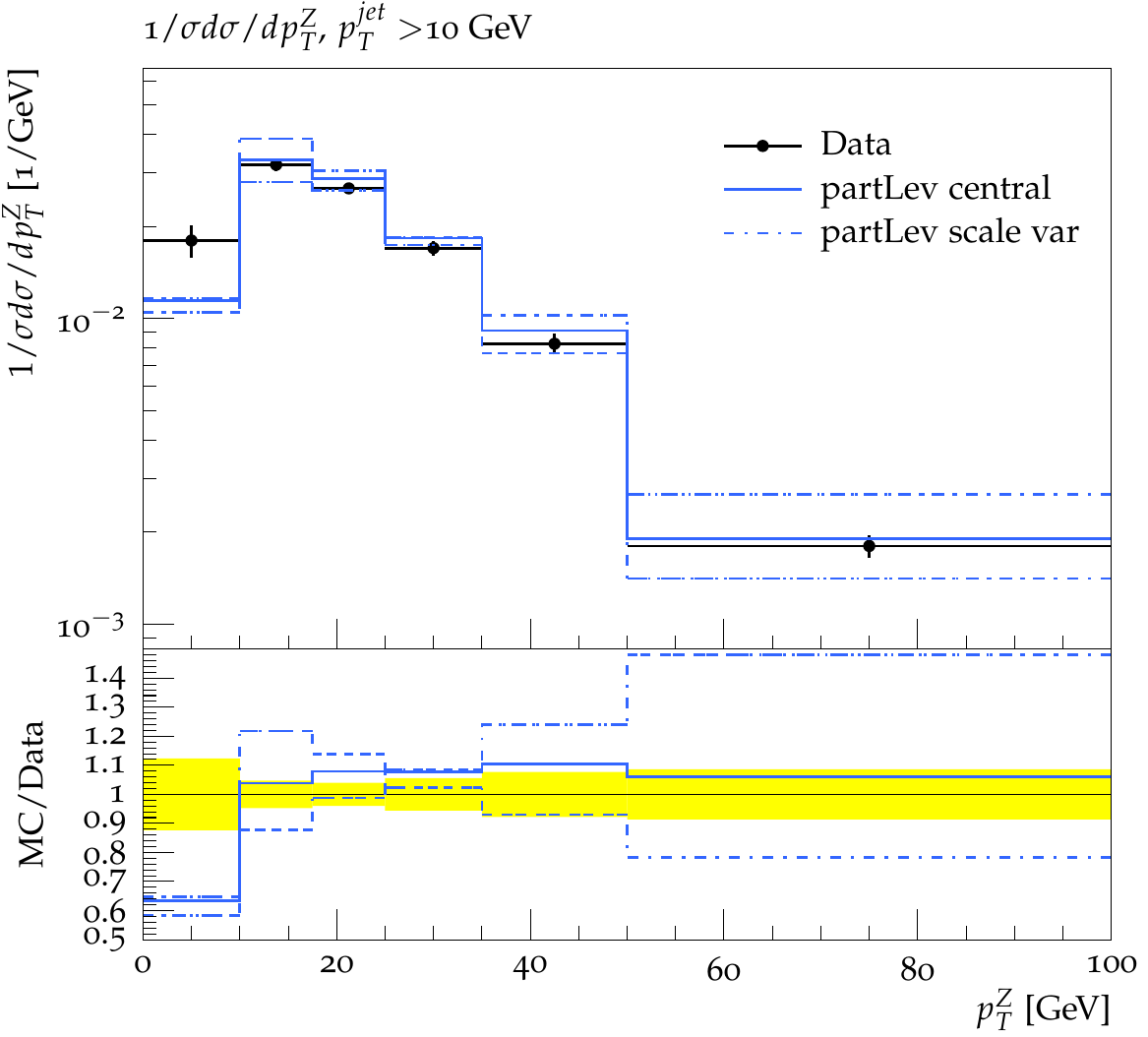}}
\hfil
\subfloat[$p_T^Z$ \normalfont{for} $p_T^{\mathrm{jet}}>20$ \normalfont{GeV}]{
\includegraphics[width=0.48\textwidth]{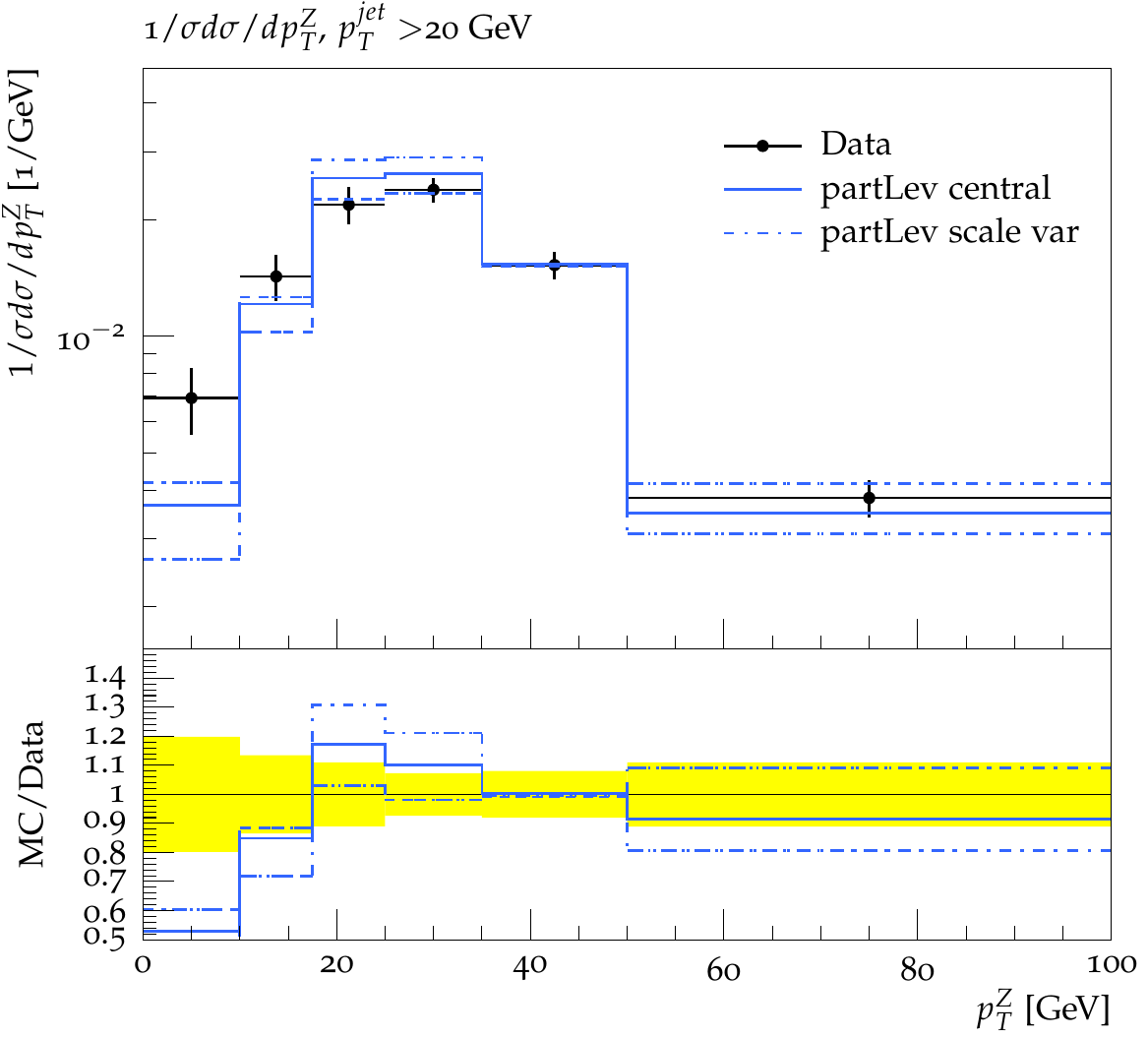}}
\caption{Scale variation for $\Delta\phi$ and $p_T^Z$ distributions calculated at the parton
level (without any showers) using a factorization/renormalization scale $\mu=\sqrt{m_Z^2+(p_T^{\rm jet})^2}$. 
Compared to the LHCb measurements~\cite{Aaij:2013nxa}.}
\label{fig:partLev_scaleUncer}
\end{figure*}
%----------------
In Fig.~\ref{fig:partLev_scaleUncer} we show distributions of $\Delta\phi$ and $p_T^Z$
calculated using two off-shell initial-state partons for two different cuts on the jet transverse momentum (as used by LHCb):
 $p_T^{\rm jet}>10$~GeV and $p_T^{\rm jet}>20$~GeV. The renormalization
and factorization scale was set to $\mu=\sqrt{m_Z^2+(p_T^{\rm jet})^2}$, which is varied by a factor of two up and down in order to estimate the 
scale uncertainty.
We can see that  for both cut choices the description of the $\Delta\phi$ distribution is very
good and the data lie within the scale uncertainties. The measurement of  $p_T^Z$ is also  well described, with the exception of the lowest $p_T$-region, where the data are underestimated.
This is the region that is driven by the initial state transverse
momentum described by  the TMDs. In a collinear parton-level $2\to2$ calculation the region below the $p_T^{\rm jet}$ cut would be empty.
%

%----------------
\begin{figure*}[!htb]
\centering{}
\subfloat[$\Delta\phi$ \normalfont{for} $p_T^{\mathrm{jet}}>10$ \normalfont{GeV}]{
\includegraphics[width=0.48\textwidth]{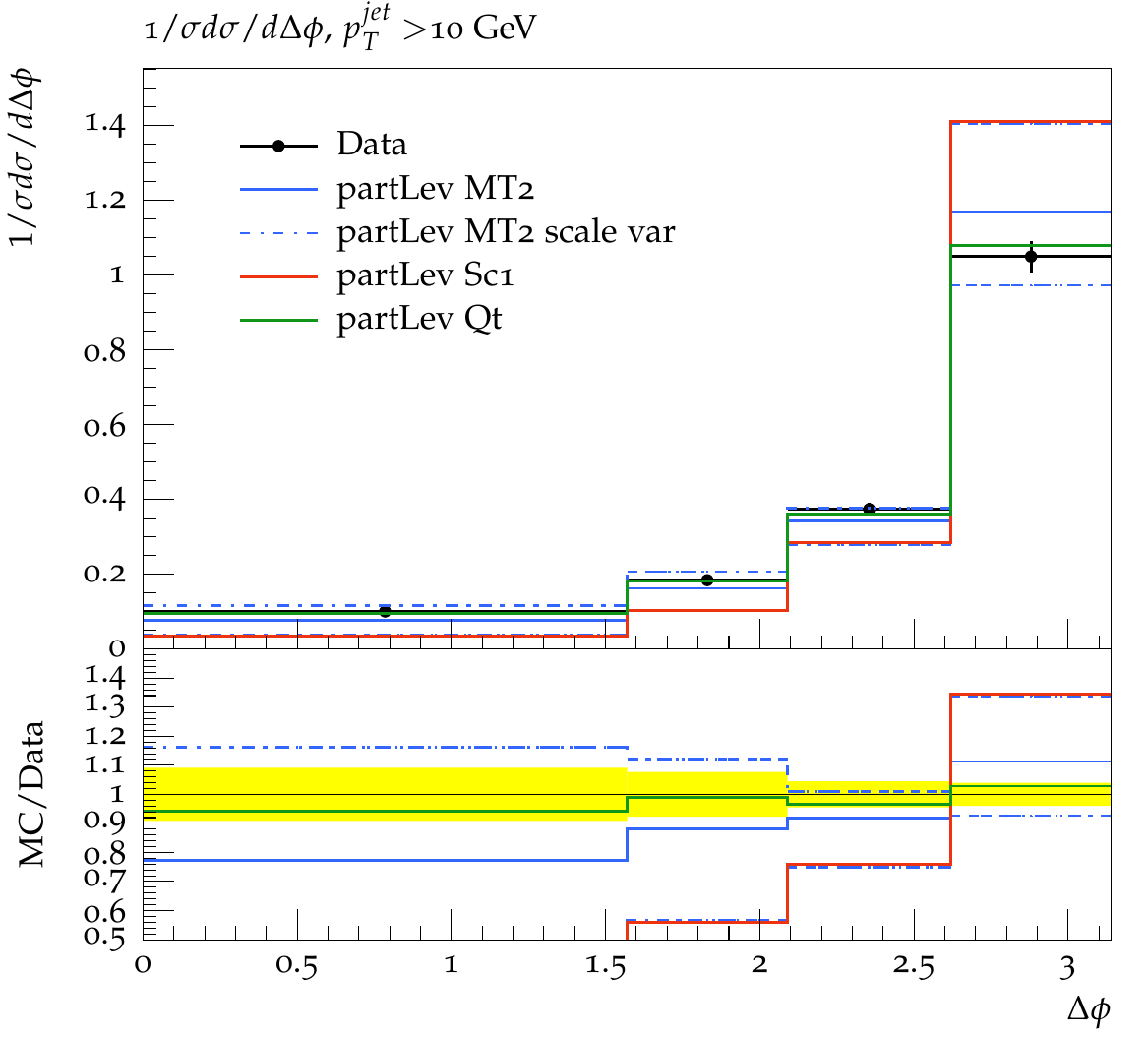}}
\hfil
\subfloat[$\Delta\phi$ \normalfont{for} $p_T^{\mathrm{jet}}>20$ \normalfont{GeV}]{
\includegraphics[width=0.48\textwidth]{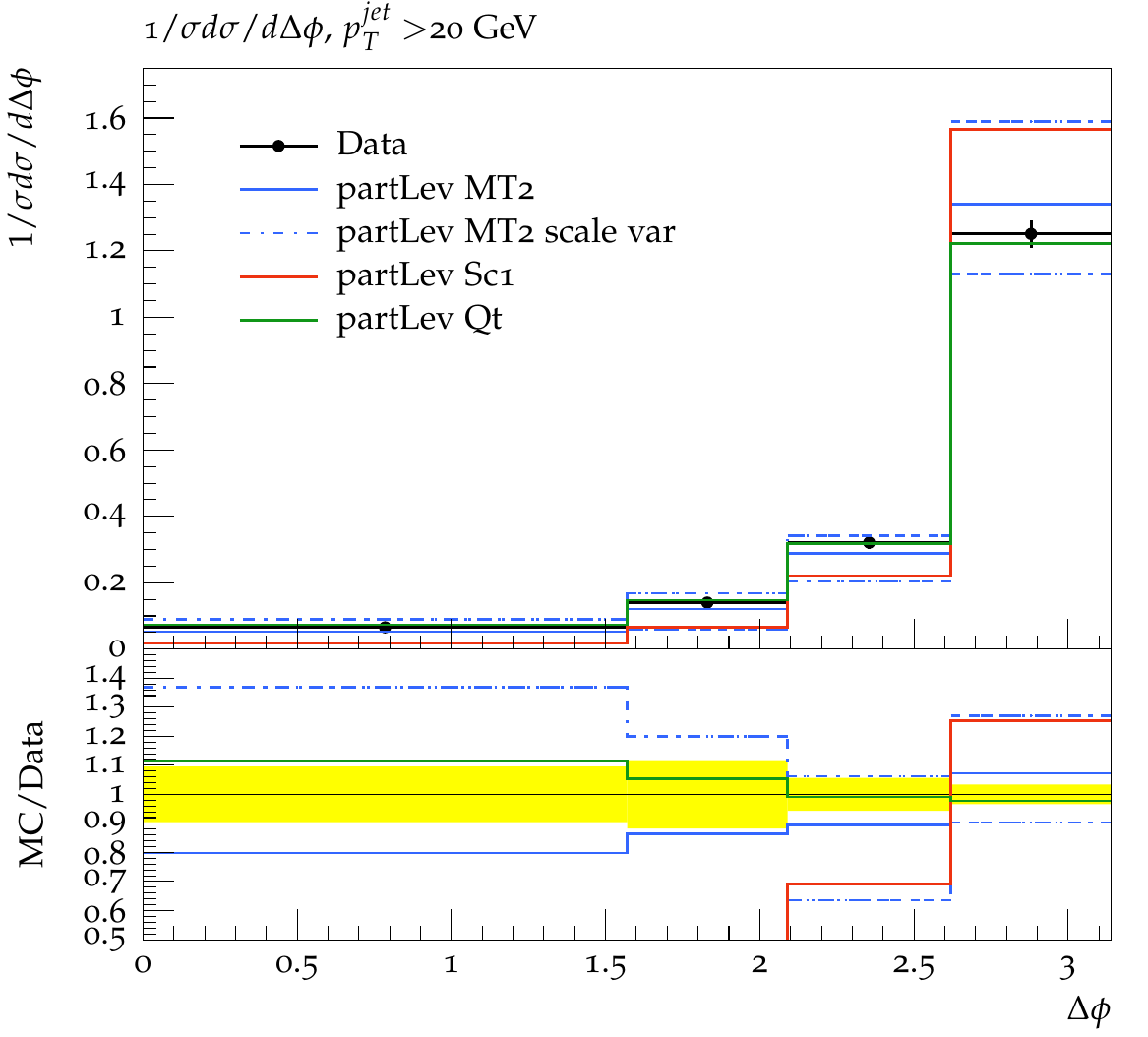}}
\\
\subfloat[$p_T^Z$ \normalfont{for} $p_T^{\mathrm{jet}}>10$ \normalfont{GeV}]{
\includegraphics[width=0.48\textwidth]{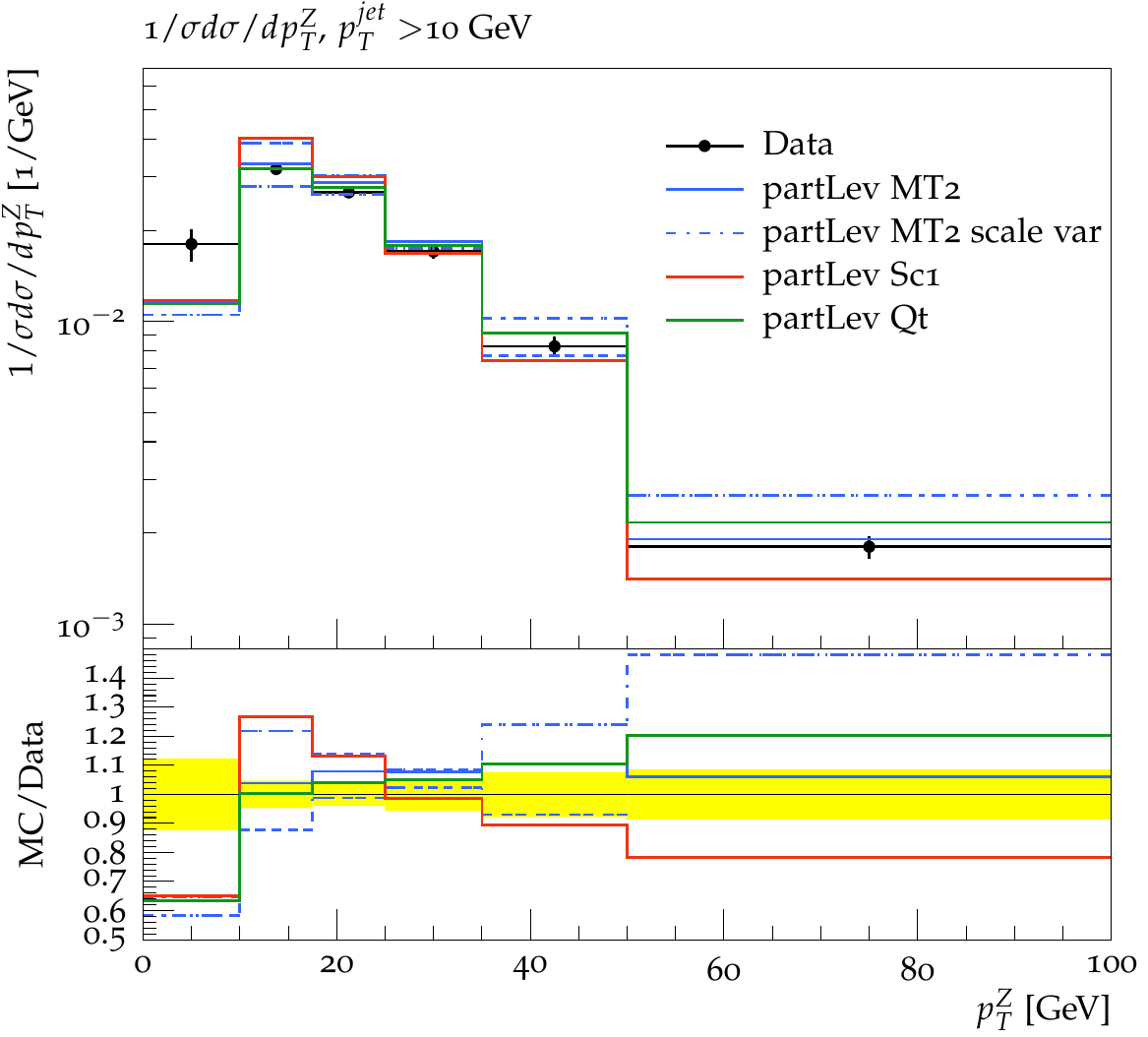}}
\hfil
\subfloat[$p_T^Z$ \normalfont{for} $p_T^{\mathrm{jet}}>20$ \normalfont{GeV}]{
\includegraphics[width=0.48\textwidth]{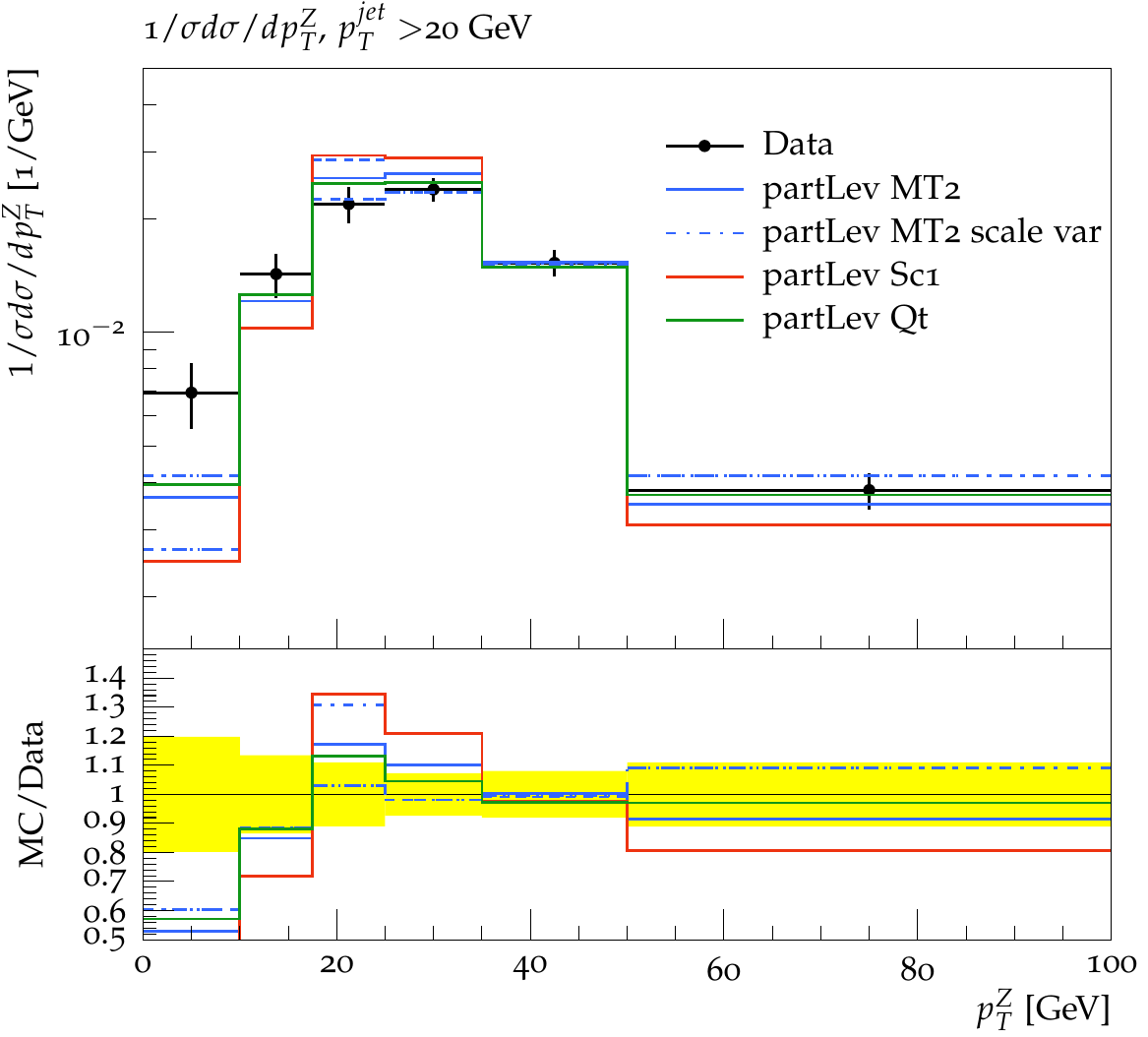}}
\caption{Comparison of different scale choices:
MT2: $\mu=\sqrt{m_Z^2+(p_T^{\rm jet})^2}$,
Sc1: $(p_T^{\rm jet}+p_T^Z+m_Z)/3$,
Qt: $\sqrt{\hat{s}+Q_t^2}$ %(which is equal to $\sqrt{m_{\rm IS}^2+p_{T\,\rm IS}^2} = \sqrt{m_{\rm FS}^2+p_{T\,\rm FS}^2}$)
for $\Delta\phi$ and $p_T^Z$ distributions calculated at the parton level (without any showers). 
Compared to the LHCb measurements~\cite{Aaij:2013nxa}.}
\label{fig:partLev_diffScale}
\end{figure*}
%----------------
%
In Fig.~\ref{fig:partLev_diffScale} we investigate the importance of the scale choice  
by comparing the calculation with $\mu=\sqrt{m_Z^2+(p_T^{\rm jet})^2}$  (Fig.~\ref{fig:partLev_scaleUncer}) with 
different choices: $\mu=(p_T^{\rm jet}+p_T^Z+m_Z)/3$
and $\mu=\sqrt{\hat{s}+Q_t^2}$ with $Q_t=p_T^{\mathrm{initial\, state}}$.
We can see that the scale $\mu=\sqrt{\hat{s}+Q_t^2}$, which is motivated by angular ordering \cite{Jung:2003wu}, gives results similar to our original scale choice, while the choice $\mu=(p_T^{\rm jet}+p_T^Z+m_Z)/3$ leads to significant differences. However,  
most of the time the results are still within the scale uncertainty of the results from Fig.~\ref{fig:partLev_scaleUncer}. 
Based on this observation, we will use the scale $\mu=\sqrt{m_Z^2+(p_T^{\rm jet})^2}$ in the following.
%

%
%----------------
\begin{figure*}[!htb]
\centering{}
\subfloat[$\Delta\phi$ \normalfont{for} $p_T^{\mathrm{jet}}>10$ \normalfont{GeV}]{
\includegraphics[width=0.48\textwidth]{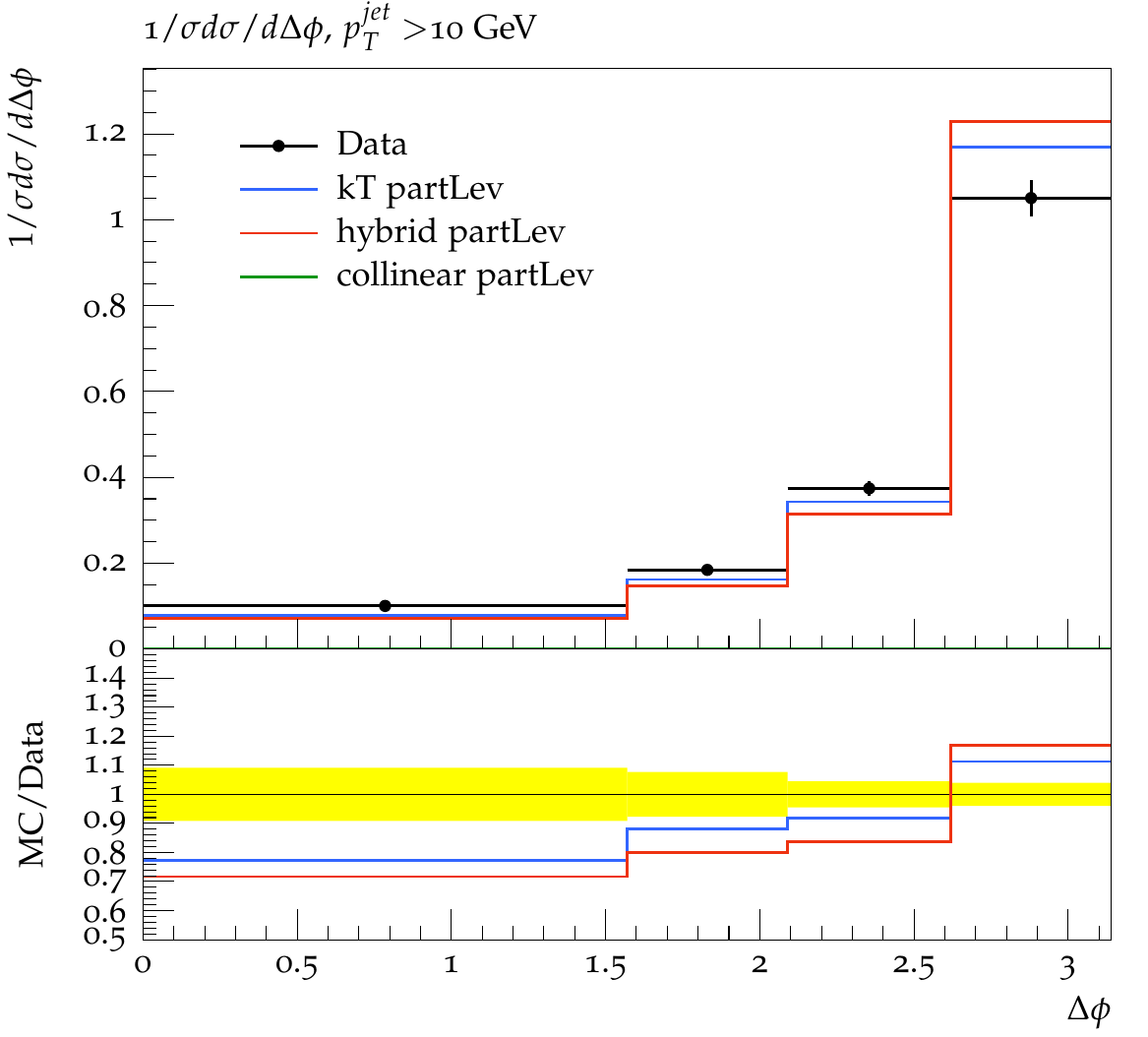}}
\hfil
\subfloat[$\Delta\phi$ \normalfont{for} $p_T^{\mathrm{jet}}>20$ \normalfont{GeV}]{
\includegraphics[width=0.48\textwidth]{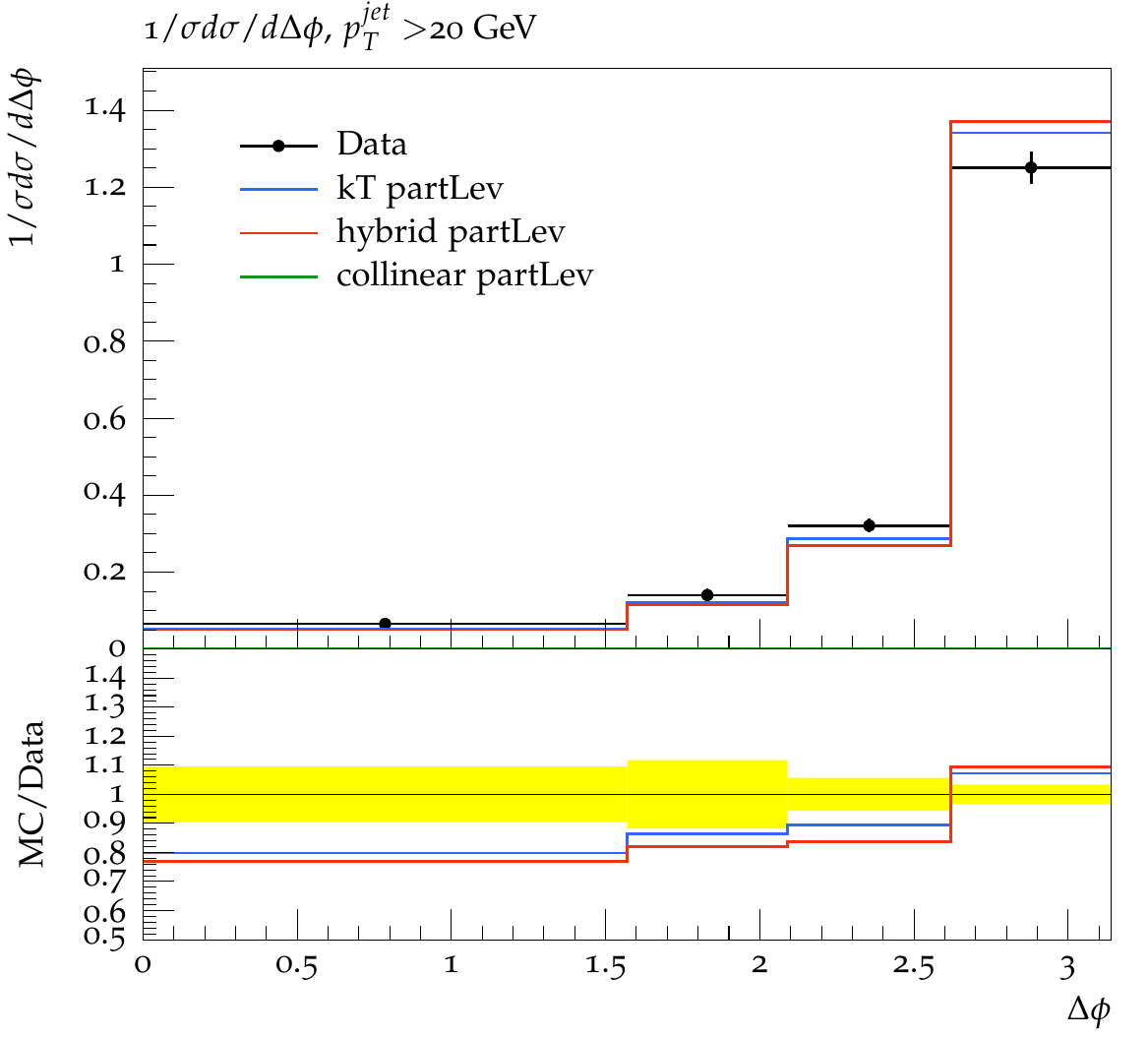}}
\\
\subfloat[$p_T^Z$ \normalfont{for} $p_T^{\mathrm{jet}}>10$ \normalfont{GeV}]{
\includegraphics[width=0.48\textwidth]{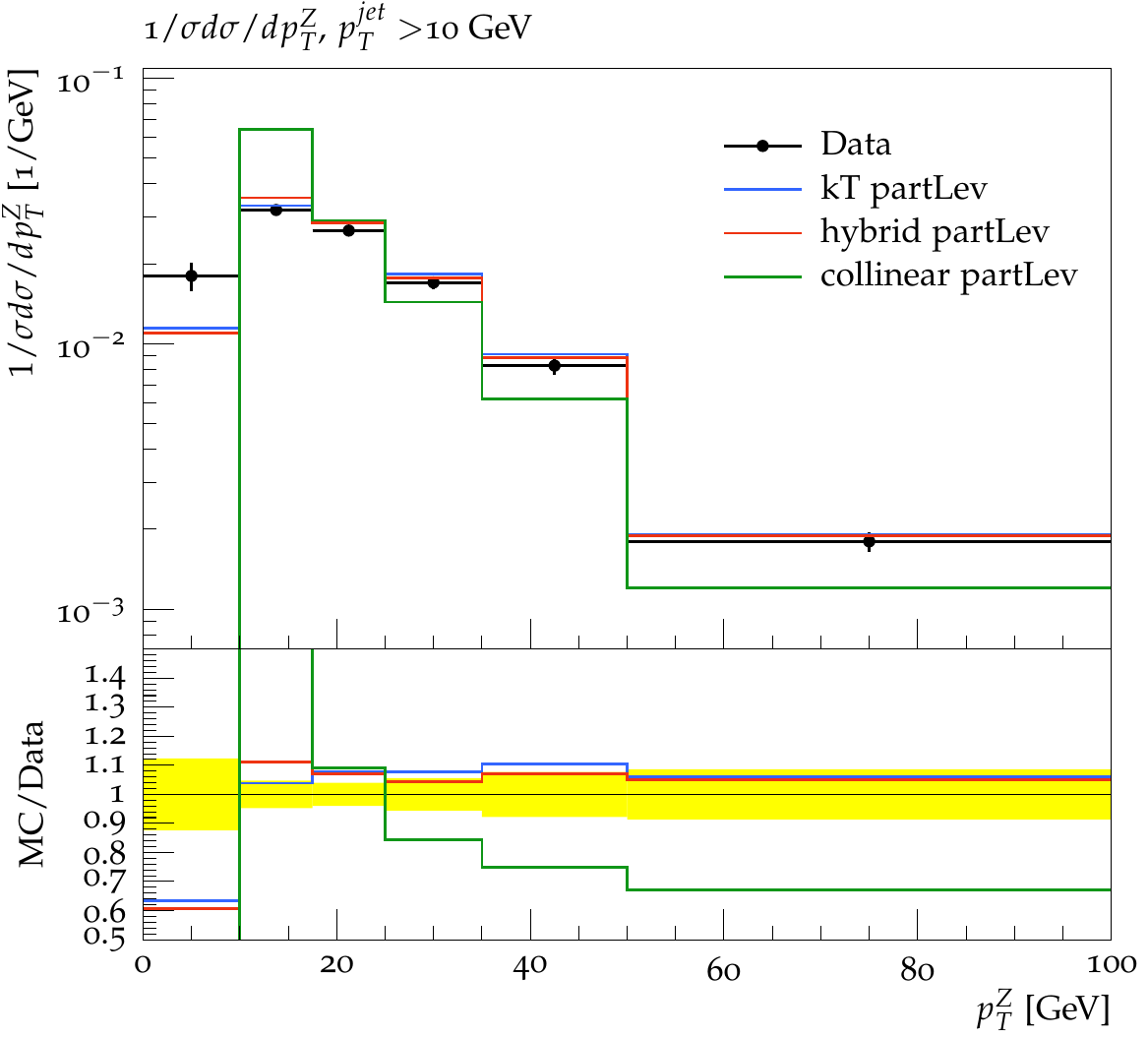}}
\hfil
\subfloat[$p_T^Z$ \normalfont{for} $p_T^{\mathrm{jet}}>20$ \normalfont{GeV}]{
\label{subfig:partLev_kT-hyb-coll_pTZ10}
\includegraphics[width=0.48\textwidth]{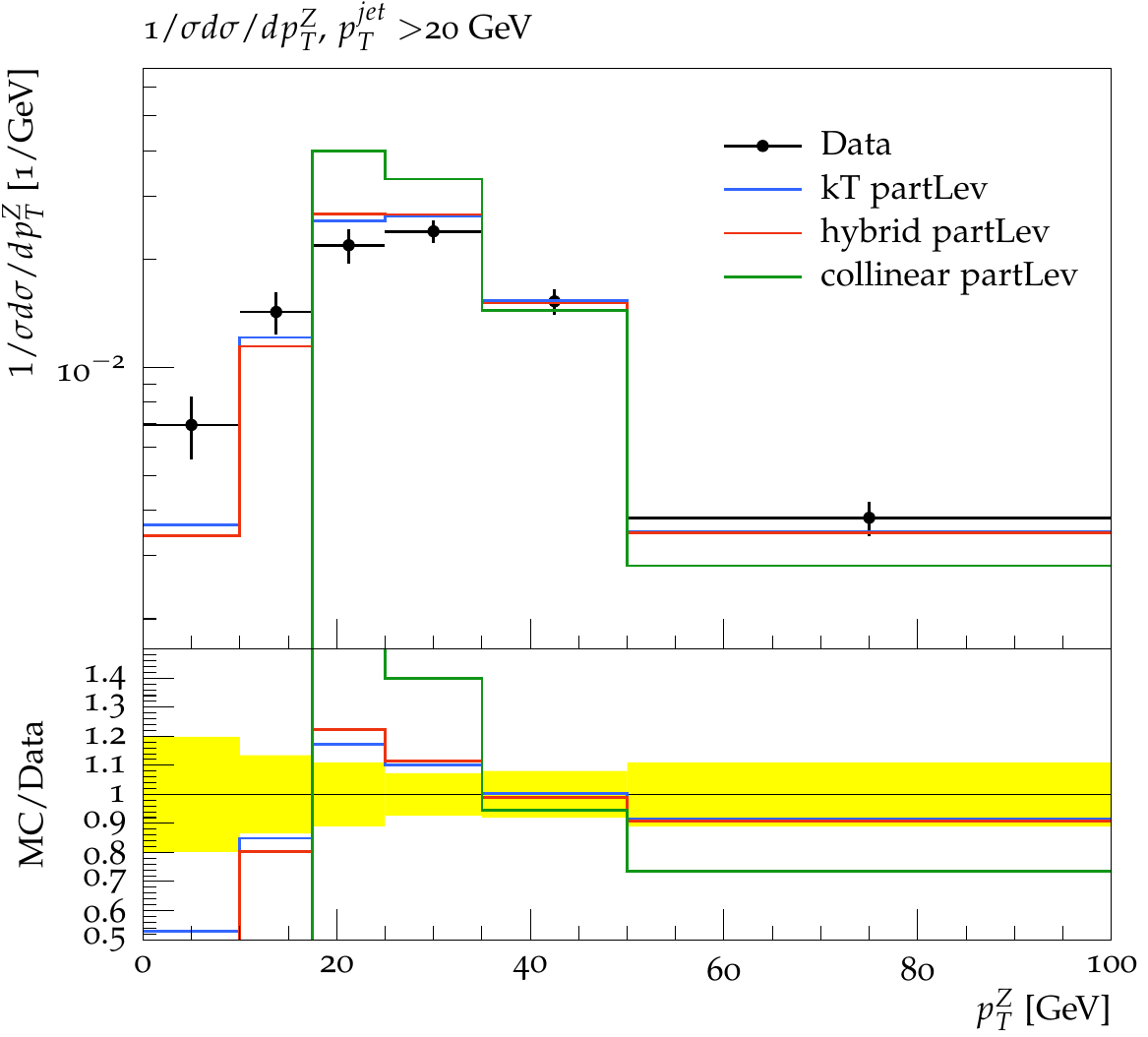}}
\caption{Comparison of $\Delta\phi$ and $p_T^Z$ distributions calculated on parton
level (without any showers) and the LHCb measurements~\cite{Aaij:2013nxa}. The factorization/renormalization scale
$\mu=\sqrt{m_Z^2+(p_T^{\rm jet})^2}$ is used for:
$k_T$-factorization (two  off-shell initial partons),
hybrid factorization (one off-shell and one on-shell initial parton),
and collinear factorization (two on-shell partons).}
\label{fig:partLev_kT-hyb-coll}
\end{figure*}
%----------------
%
In Fig.~\ref{fig:partLev_kT-hyb-coll} we compare the predictions obtained within $k_T$-factorization
(presented in Fig.~\ref{fig:partLev_scaleUncer}) to the corresponding predictions obtained in the
hybrid (the lower $x$ initial state parton is off-shell while the higher $x$ initial state parton is on-shell) and collinear approaches at parton level. 
We can see that for both $\Delta\phi$ and $p_T^Z$ distributions (with both cut choices) the results of the hybrid approach are very similar to the ones of $k_T$-factorization. 
This shows that the hybrid factorization works well where it is expected to work, i.e.\ in the forward region where asymmetric values of $x_1$ and $x_2$ are probed. 
The key point in this study are the TMDs and collinear PDFs used: the $k_T$-integrated PB-TMD is identical to the collinear PDF by construction. When 
HE-factorization is used in the forward rapidity region, the $k_T$ on the large $x$ side is limited and the matrix elements with two off-shell initial state particles 
effectively become on-shell--off-shell, and in the process of evaluating the cross section the PDF on that side is integrated to yield a collinear PDF.
In the hybrid factorization calculation, the collinear PDF was used from the beginning. 
A more detailed analysis of it will be provided in Sec.~\ref{sec:correlations}
where we discuss correlations of transverse momenta and $x$ values of the initial partons.

In Fig.~\ref{fig:partLev_kT-hyb-coll} we additionally present results obtained within collinear factorization. 
The motivation is to show that a purely collinear approach is not able to describe
certain distributions at leading order (LO) and higher order corrections are needed. 
This is especially visible for the $\Delta\phi$ distribution which reduces to a delta function at LO. 
The cut on the $p_T^{\mathrm{jet}}$ restricts also the $p_T^Z$ distribution since 
there is no recoil that would allow the $Z$ boson to gain additional transverse momentum. Later we will compare our results also with predictions
in collinear factorization at LO but including parton showers as well as with next-to-leading order (NLO) predictions.

% \clearpage
%%%%%%%%%%%%%%%
\subsection{Showered results}
%%%%%%%%%%%%%%%
%----------------
\begin{figure*}[!htb]
\centering{}
\subfloat[$\Delta\phi$ \normalfont{for} $p_T^{\mathrm{jet}}>10$ \normalfont{GeV}]{
\includegraphics[width=0.48\textwidth]{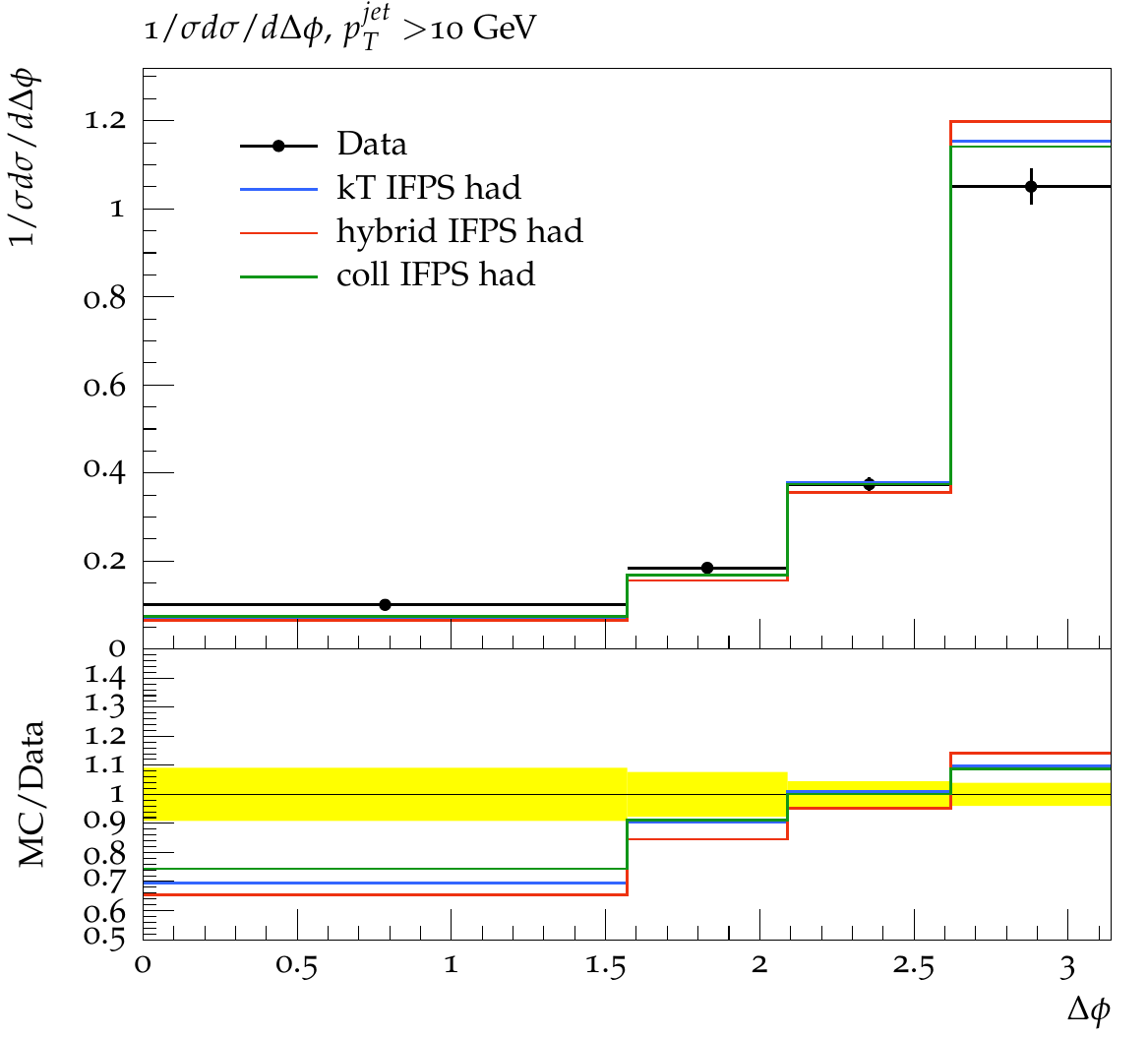}}
\hfil
\subfloat[$\Delta\phi$ \normalfont{for} $p_T^{\mathrm{jet}}>20$ \normalfont{GeV}]{
\includegraphics[width=0.48\textwidth]{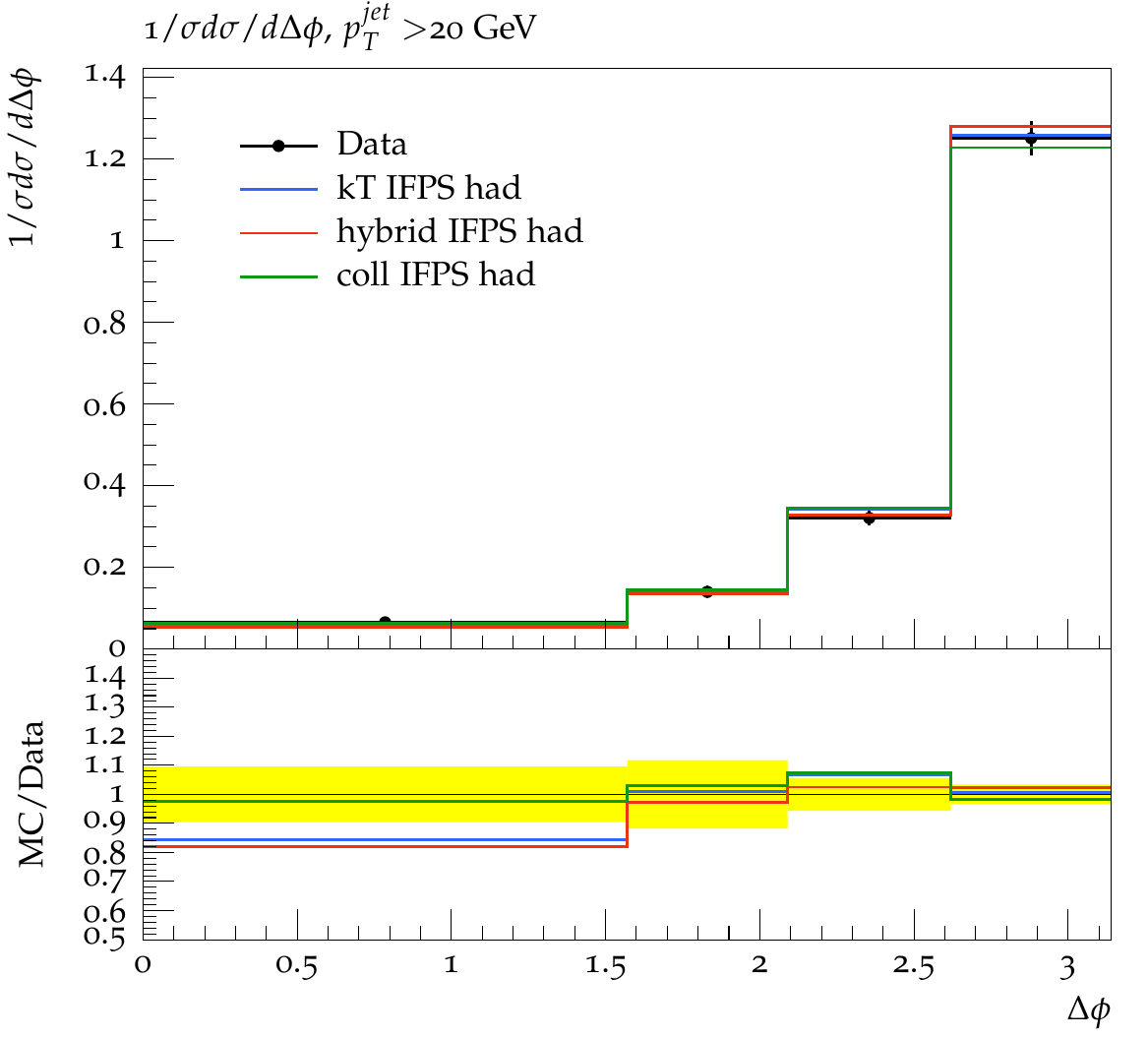}}
\\
\subfloat[$p_T^Z$ \normalfont{for} $p_T^{\mathrm{jet}}>10$ \normalfont{GeV}]{
\includegraphics[width=0.48\textwidth]{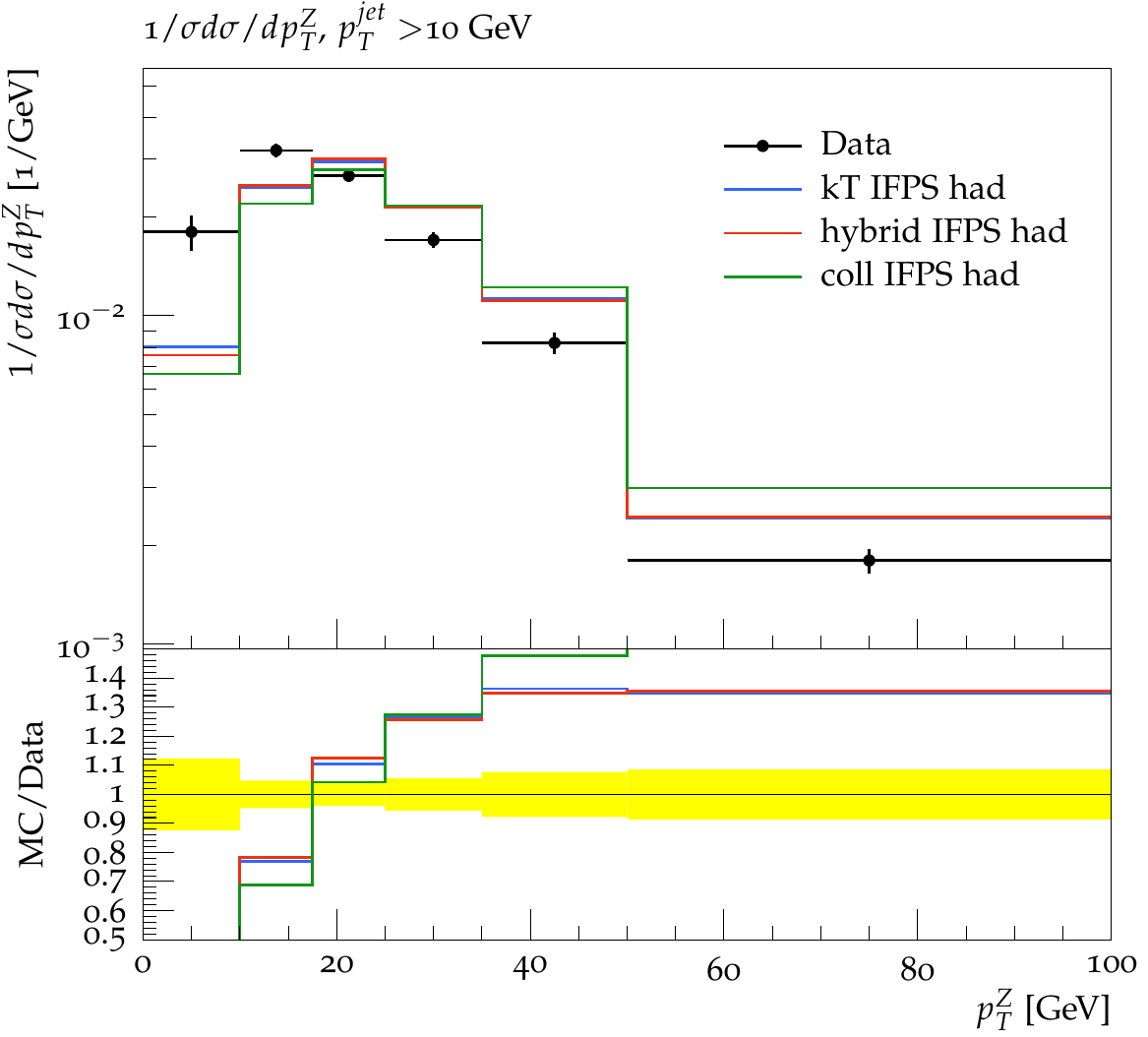}}
\hfil
\subfloat[$p_T^Z$ \normalfont{for} $p_T^{\mathrm{jet}}>20$ \normalfont{GeV}]{
\label{subfig:IFPShad_kT-hyb-coll_pTZ10}
\includegraphics[width=0.48\textwidth]{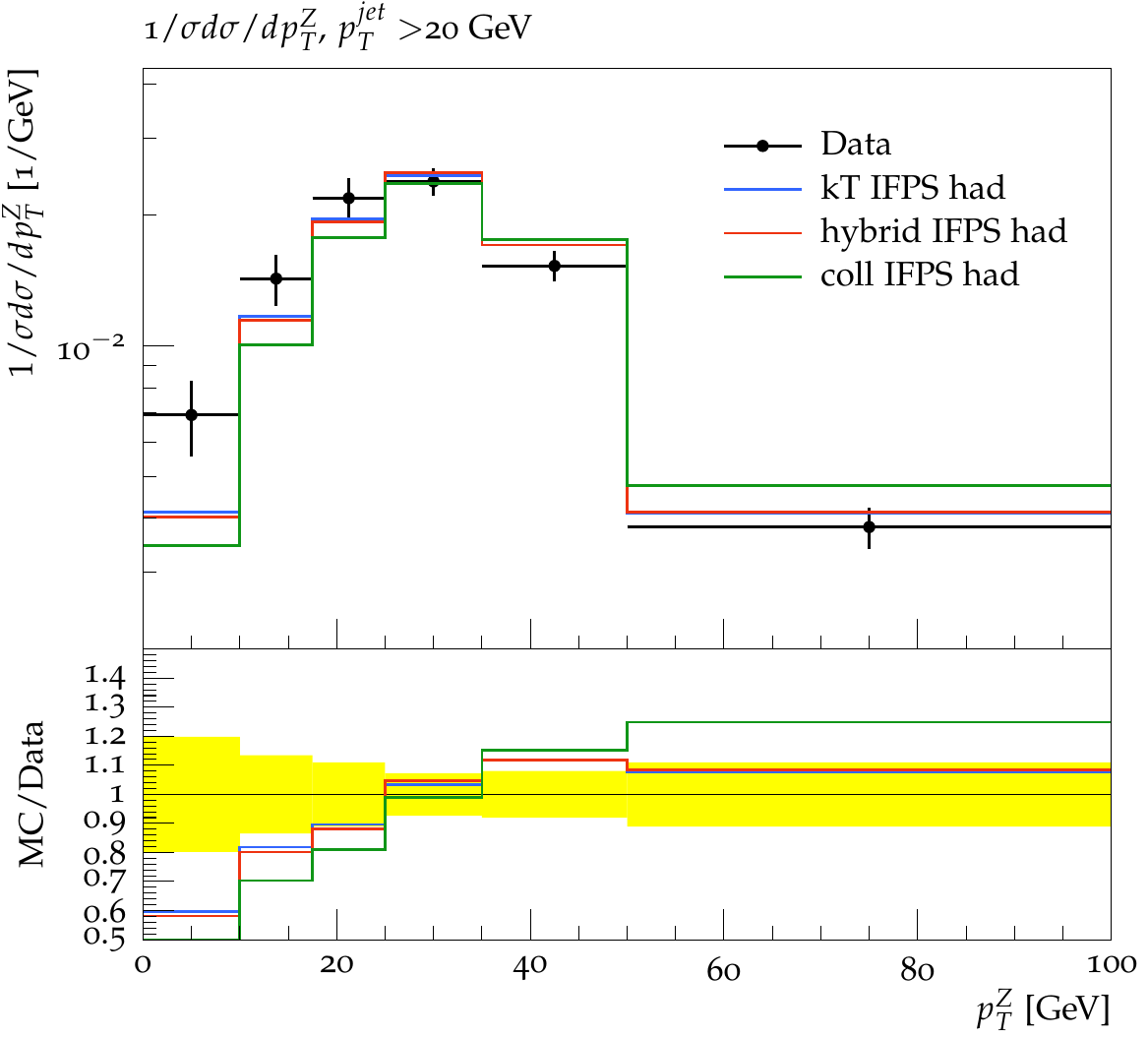}}
\caption{Comparison of $\Delta\phi$ and $p_T^Z$ distributions calculated including initial and final
state showers as well as hadronization, using factorization/renormalization scale
$\mu=\sqrt{m_Z^2+(p_T^{\rm jet})^2}$ within:
$k_T$-factorization (two off-shell initial partons),
hybrid factorization (one off-shell and one on-shell initial parton),
and collinear factorization (two on-shell partons).
Compared to the LHCb measurements~\cite{Aaij:2013nxa}.}
\label{fig:IFPShad_kT-hyb-coll}
\end{figure*}
%----------------
In this section we present results using an updated version of  the CASCADE Monte Carlo event
generator \cite{Jung:2010si,Jung:2001hx,Bury:2017jxo}, allowing to process LHE Les Houches event
files~\cite{Alwall:2006yp}, generated by \katie\ and to provide initial state parton showers
according to the PB-TMDs. The calcualtions are also supplemented with standard final state parton
shower and hadronization \cite{Sjostrand:2006za}, the multi-parton interactions are not included.
When off-shell matrix elements are used, the transverse momenta of the initial state partons are already included according to the TMDs. In case of  collinear matrix elements, first a transverse momentum is generated according to the TMD and added to the event record in such a way, that the mass of the hard process $\hat s$ is preserved, and  the event is showered, while the parton shower does not change the kinematics of the hard process (after the transverse momentum is included). In hybrid factorisation, the initial state parton shower is included only for the off-shell leg, while the other leg stays at $k_T=0$. 
It is important to stress, that the parton densities used in the calculations are applied in a consistent way.
% Note that in Fig.~\ref{fig:IFPShad_kT-hyb-coll} the $\Delta\phi$ distribution
% for the collinear calculation is fully generated by the initial state shower
% (using PB PDFs) and it is doing a rather good job describing the data.

%----------------
\begin{figure*}[!hb]
\centering{}
\subfloat[$\Delta\phi$ \normalfont{for} $p_T^{\mathrm{jet}}>10$ \normalfont{GeV}]{
\label{fig:IFPShad_scaleUncer_powheg_phi10}
\includegraphics[width=0.48\textwidth]{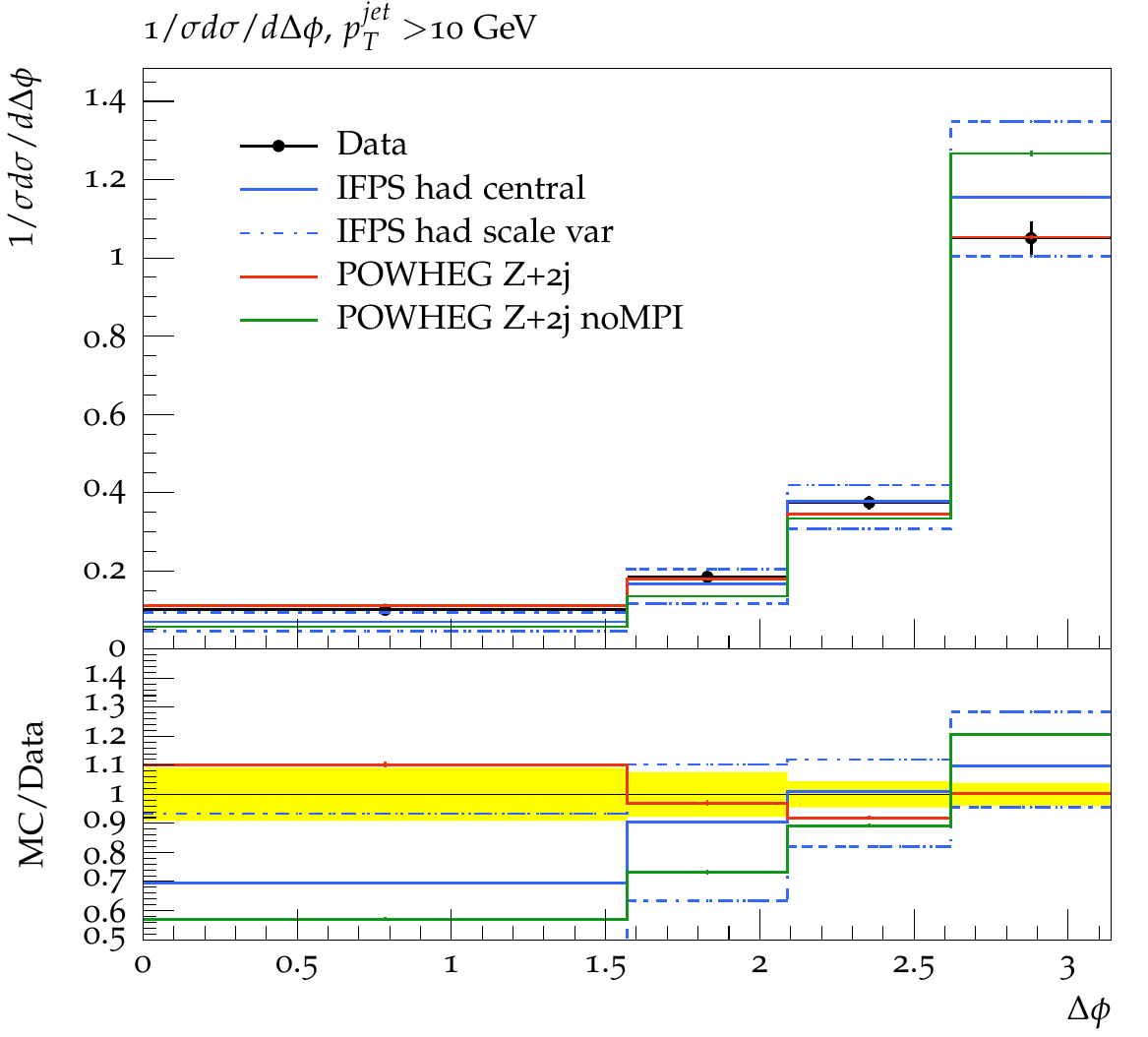}}
\hfil
\subfloat[$\Delta\phi$ \normalfont{for} $p_T^{\mathrm{jet}}>20$ \normalfont{GeV}]{
\label{fig:IFPShad_scaleUncer_powheg_phi20}
\includegraphics[width=0.48\textwidth]{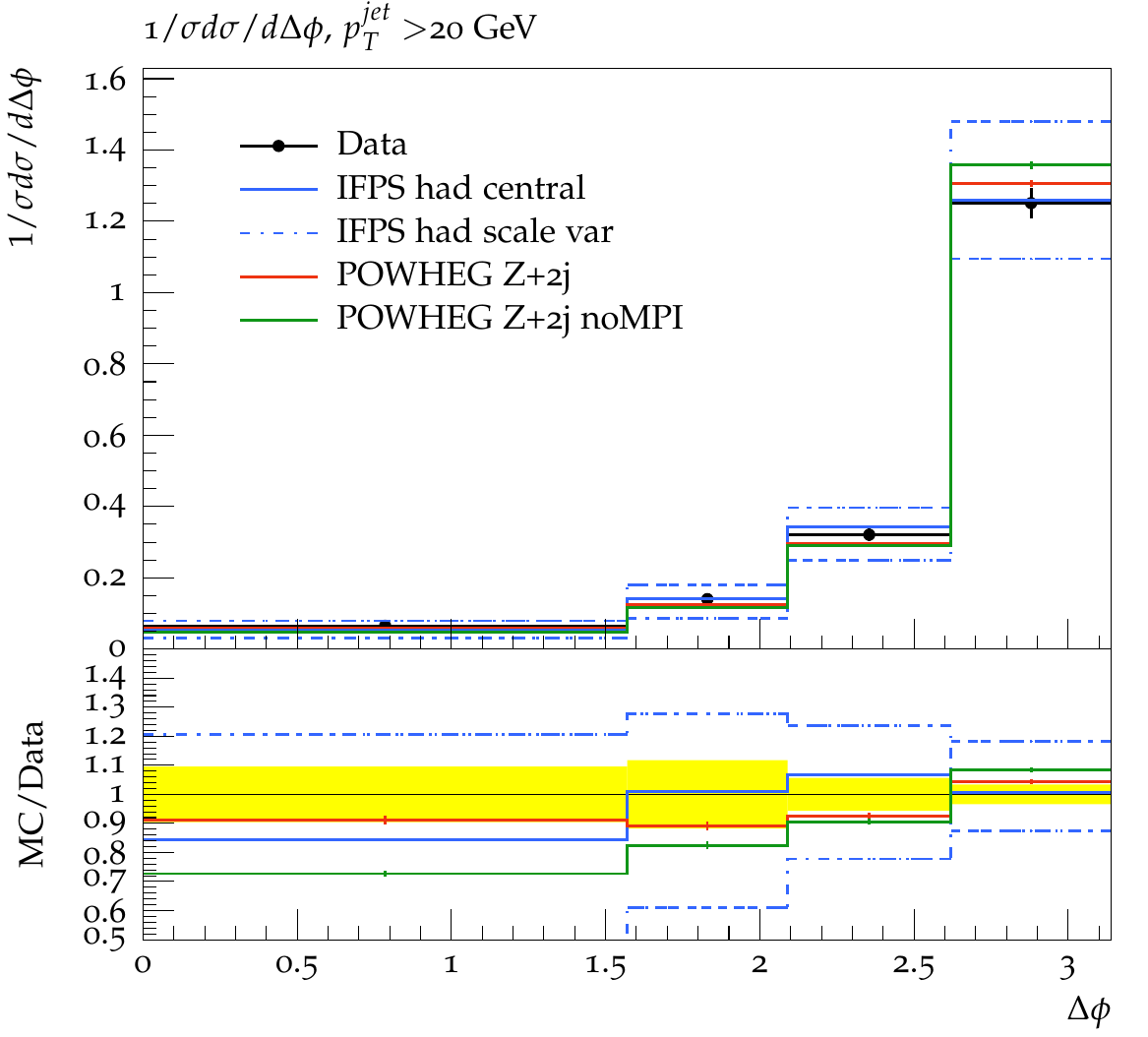}}
\\
\subfloat[$p_T^Z$ \normalfont{for} $p_T^{\mathrm{jet}}>10$ \normalfont{GeV}]{
\label{fig:IFPShad_scaleUncer_powheg_pTZ10}
\includegraphics[width=0.48\textwidth]{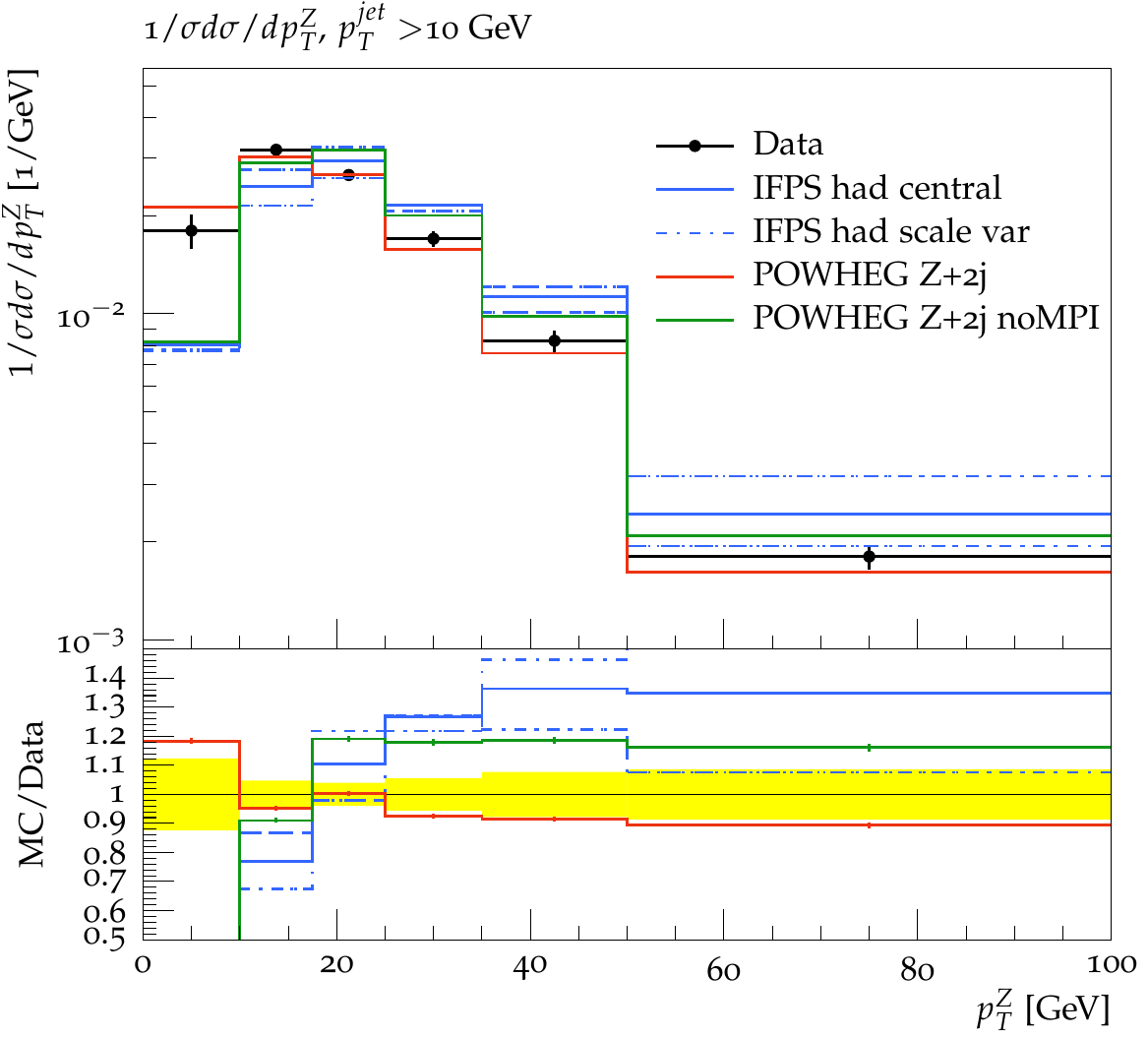}}
\hfil
\subfloat[$p_T^Z$ \normalfont{for} $p_T^{\mathrm{jet}}>20$ \normalfont{GeV}]{
\label{fig:IFPShad_scaleUncer_powheg_pTZ20}
\includegraphics[width=0.48\textwidth]{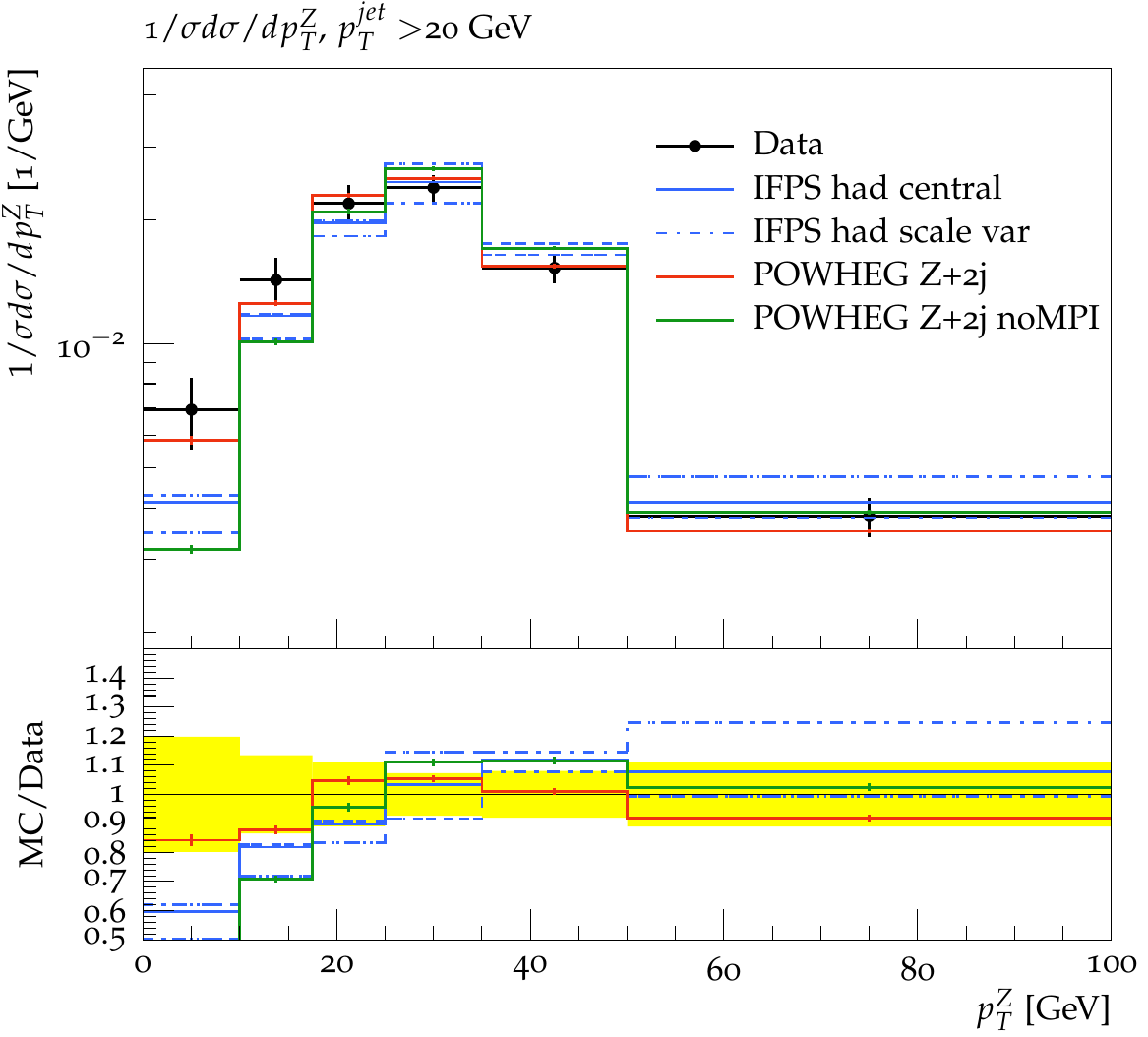}}
\caption{Scale variation for $\Delta\phi$ and $p_T^Z$ distributions calculated including
initial and final state showers as well as hadronization using factorization/renormalization
scale $\mu=\sqrt{m_Z^2+(p_T^{\rm jet})^2}$ compared with POWHEG results for $Z+2\mathrm{jet}$ production. 
Compared to the LHCb measurements~\cite{Aaij:2013nxa}.}
\label{fig:IFPShad_scaleUncer_powheg}
\end{figure*}
%----------------

In Fig.~\ref{fig:IFPShad_kT-hyb-coll} we compare results obtained using $k_T$-factorization, 
hybrid factorization and collinear factorization (using the scale
$\mu=\sqrt{m_Z^2+(p_T^{\rm jet})^2}$). We can see that, similarly to the results at the parton level, 
the $k_T$-factorization and hybrid factorization approaches give very similar results. 
On the other hand, we see a substantial change in the case of predictions from collinear factorization, 
which now is also quite similar to the other two predictions:
the $\Delta\phi$ distribution in the collinear case is no longer zero and it gives a rather good description of the data.
Note here that the $\Delta\phi$ distribution is fully generated by the TMDs (and the corresponding TMD shower). From a kinematic point of view this mimics
the features of the off-shell calculation, where the initial transverse
momenta are included from the beginning. It also points to the observation that initial state transverse momenta
are perhaps more important than the off-shellness in the matrix elements: the off-shell matrix elements would contribute a dynamical correction, while the kinematics are driven by including the transverse momenta from the TMDs.
%\comment{As suggested by Michal, we could discuss how we reach this conclusion}
While calculations obtained in HE- and hybrid  factorization give essentially the same results, a difference to results from collinear factorization with parton shower is observed. The parton densities and the parton showers are the same for all calculations, and the difference comes entirely from the different matrix elements used (off-shell versus on-shell).

The $\Delta\phi$ distributions are rather well described by all approaches, while in the $p_T^Z$ distribution, all 
approaches predict a too small cross section at small $p_T^Z$,  the high-energy and hybrid factorization approach are 
slightly better  than the collinear calculation with TMD showers included. 
%Interestingly, including showers and hadronization degrades of the description 
%of the data by the high-energy and hybrid approach compared to the parton level predictions
%(compare Fig.~\ref{subfig:IFPShad_kT-hyb-coll_pTZ10} and Fig.~\ref{subfig:partLev_kT-hyb-coll_pTZ10}).
%\comment{Is this really true ?}
We have checked that changing the scale $\mu$ gives similar effects as observed without TMD showers.

Finally, in Fig.~\ref{fig:IFPShad_scaleUncer_powheg} we compare  $k_T$-factorization predictions with parton showers and hadronization (including scale variations by a factor of two)
%obtained with our preferred scale choice $\mu=\sqrt{m_Z^2+(p_T^{\rm jet})^2}$
with a collinear NLO calculation of $Z$-boson and 2 jets performed in POWHEG
(with MiNLO method~\cite{Hamilton:2012np,Hamilton:2012rf}), using the HERAPDF20\_NLO collinear
PDFs~\cite{Abramowicz:2015mha} and the \pythia 8 tune CUETP8M1~\cite{Khachatryan:2015pea}. We show predictions with and without multi-parton interactions (MPI).
The measured $\Delta\phi$ distribution is well described by the NLO collinear calculation (Figs.~\ref{fig:IFPShad_scaleUncer_powheg_phi10} and
\ref{fig:IFPShad_scaleUncer_powheg_phi20}), when MPI is included.
Similarly, the POWHEG calculation agrees rather well with the measured 
 $p_T^Z$ distribution (Figs.~\ref{fig:IFPShad_scaleUncer_powheg_pTZ10} and
\ref{fig:IFPShad_scaleUncer_powheg_pTZ20}), if MPI is included, 
in particular at small $p_T^Z$, below the $p_T^{\mathrm{jet}}$ cut-off. 
It is interesting to note, that at low  $p_T^Z$  the description of the measurement is significantly improved
when MPI is included, meaning that some of the low $p_T$ jets come from MPI.
If the contribution from MPI is switched off, then the predictions
calculated with collinear NLO $Z+2$ jet and with off-shell matrix elements in LO $k_T$-factorization
agree rather well. This observation confirms that the distributions using off-shell matrix elements are similar to the ones obtained by a collinear NLO calculation.

% the POWHEG calculation agrees rather well at small $p_T^Z$, below the $p_T^{\mathrm{jet}}$
% cut-off,  with the measurements. It is remarkable how similar the distributions are when
% calculated with collinear NLO $Z$+ 2 jet and off-shell matrix elements, once the parton
% densities and $\alpha_s$ are used consistently. The difference in description 
% at small $p_T^Z$ comes mainly from the POWHEG internal Sudakov form factor.

%The higher order corrections seem more important for the $p_T^Z$ distribution which is
% now better described by the collinear NLO computations. This is especially important
% for the lowest $p_T^Z$ bin which is below the $p_T^{\mathrm{jet}}>10(20)$ GeV cut-off
% where additional emission (extra jet) allows for additional recoil. 
%

The differential distributions measured by LHCb are normalized to the total cross-section
which means we could not judge the normalization of our predictions.
In order to do it in Table~\ref{tab:totXS} we show a comparison of the measured total
cross-section with the different predictions.
% In Table~\ref{tab:totXS} we show a comparison of the measurement to different predictions.
All LO predictions (using $\alpha_s(m_Z) = 0.118$) are significantly smaller than the measurements.
This is partly caused by the use of NLO PDFs and 2-loop $\alpha_s$. We have verified that using
consistently LO PDFs and strong coupling (we used HERAPDF20\_LO\_EIG PDF set) increases the cross
section and brings it closer to the measurement but it does not explains the whole difference. 
%
% \begin{table}
% \begin{center}
% %\begin{tabular}{|l|l|l|l|l|l|}
% \begin{tabular}{|c|c|c|c|c|c|}
% \hline
% $p_T^{\mathrm{jet}}$    & LHCb          & $k_T$                & hybrid               & coll. (LO) & coll. (NLO) \\
% $[\mathrm{GeV}]$        &               &                      &                      &            & with MPI \\
% \hline
% $20$ & $6.3\pm0.55$  & $3.62\pm1.1\cdot10^{-5}$ & $3.92\pm2.0\cdot10^{-5}$ & $4.63\pm3.8 \cdot10^{-5}$& $5.48\pm4.0\cdot10^{-1}$\\
% \hline
% $10$  & $16.0\pm1.36$ & $7.46\pm2.3\cdot10^{-5}$ & $7.91\pm4.1\cdot10^{-5}$ & $9.07\pm7.4\cdot10^{-5}$& $15.49\pm1.2$\\
% \hline
% \end{tabular}
% \end{center}
% \caption{Total cross-section for $Z+\mathrm{jet}$ as measured by LHCb~\cite{Aaij:2013nxa}
% and predicted by the corresponding calculations in $k_T$, hybrid and collinear factorization.
% All the calculations include showering and hadronization and were performed using the scale $\mu=\sqrt{m_Z^2+(p_T^{\rm jet})^2}$. 
% The uncertainties of the theoretical predictions are numerical integration errors.}
% \label{tab:totXS}
% \end{table}
%
\begin{table}
\begin{center}
\resizebox{\textwidth}{!}{
%\begin{tabular}{|l|l|l|l|l|l|}
\begin{tabular}{|c|c|c|c|c|c|c|}
\hline
$p_T^{\mathrm{jet}}$ & LHCb                  & $k_T$                 & hybrid                & coll.\ (LO)            & coll.\ (NLO) & coll.\ (NLO)\\
$[\mathrm{GeV}]$     &                       &                       &                       &                       & no MPI      & with MPI \\
\hline
$20$                 & $6.3\pm0.55$          & $3.6\pm1\cdot10^{-5}$ & $3.9\pm2\cdot10^{-5}$ & $4.6\pm4\cdot10^{-5}$ & $5.2\pm0.04$ & $5.5\pm0.4$\\
\hline
$10$                 & $16.0\pm1.36$         & $7.5\pm2\cdot10^{-5}$ & $7.9\pm4\cdot10^{-5}$ & $9.1\pm7\cdot10^{-5}$ & $11.4\pm0.09$ & $15.5\pm1.2$\\
\hline
\end{tabular}
}
\end{center}
\caption{Total cross-section for $Z+\mathrm{jet}$ as measured by LHCb~\cite{Aaij:2013nxa}
and predicted by the corresponding calculations in $k_T$, hybrid and collinear factorization.
All the calculations include showering and hadronization and were performed using the scale $\mu=\sqrt{m_Z^2+(p_T^{\rm jet})^2}$. 
The uncertainties of the theoretical predictions are numerical integration errors.}
\label{tab:totXS}
\end{table}

Additionally, Table~\ref{tab:totXS} shows us also the importance of the MPI for the total cross section.

% \clearpage

%%%%%%%%%%%%%%%
\subsection{Correlations}
\label{sec:correlations}
%\comment{What do we want to show in this section ?}
%%%%%%%%%%%%%%%

In this section we will analyse correlations between longitudinal and transverse components
of the initial state partons. 
%This will help us understand the structure of the results from the previous sections and the differences between high-energy and hybrid factorization.

\begin{figure*}[!htb]
\centering{}
\subfloat[$k_T$-approach]{
\includegraphics[width=0.48\textwidth]{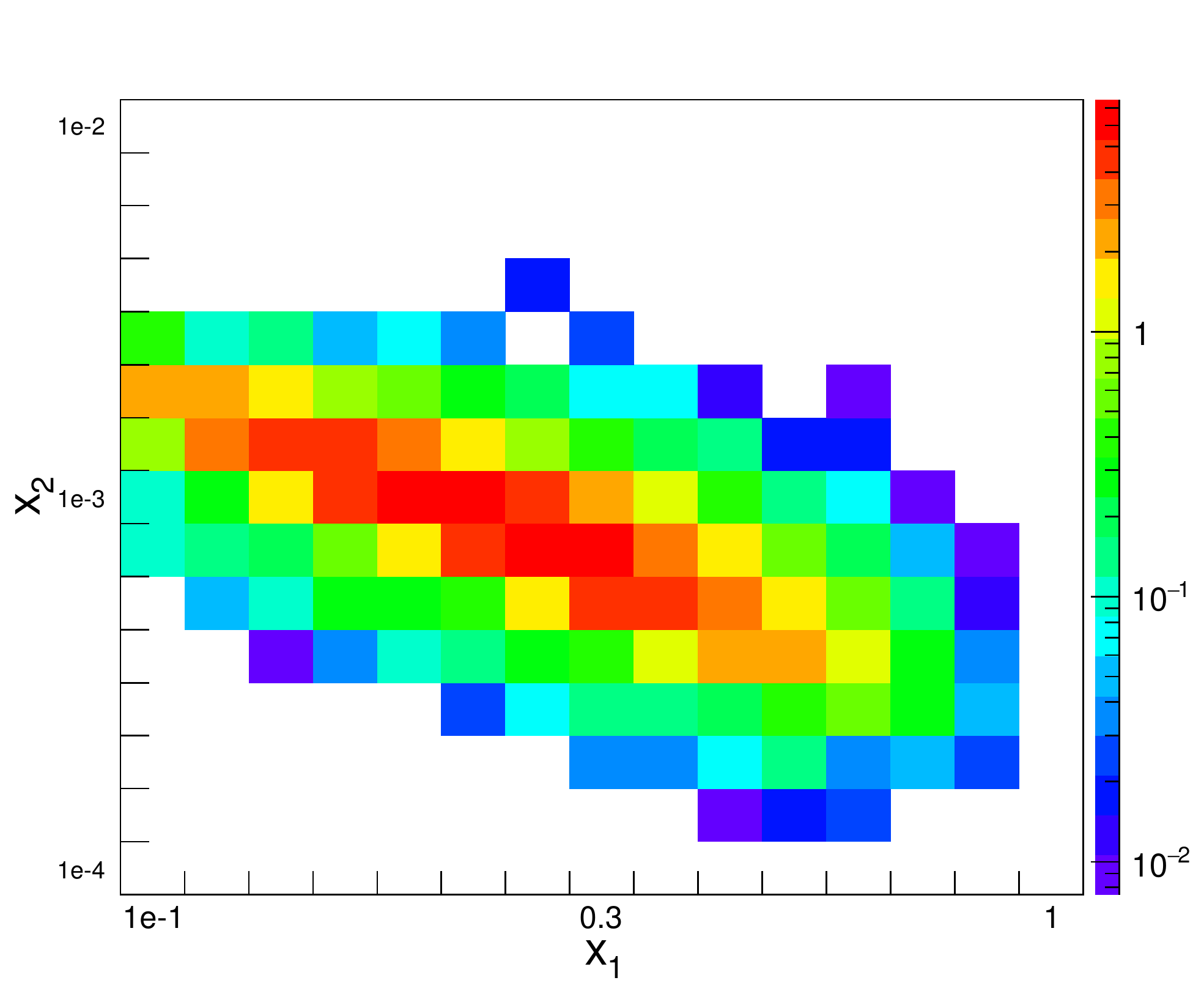}}
\hfil
\subfloat[hybrid-approach]{
\includegraphics[width=0.48\textwidth]{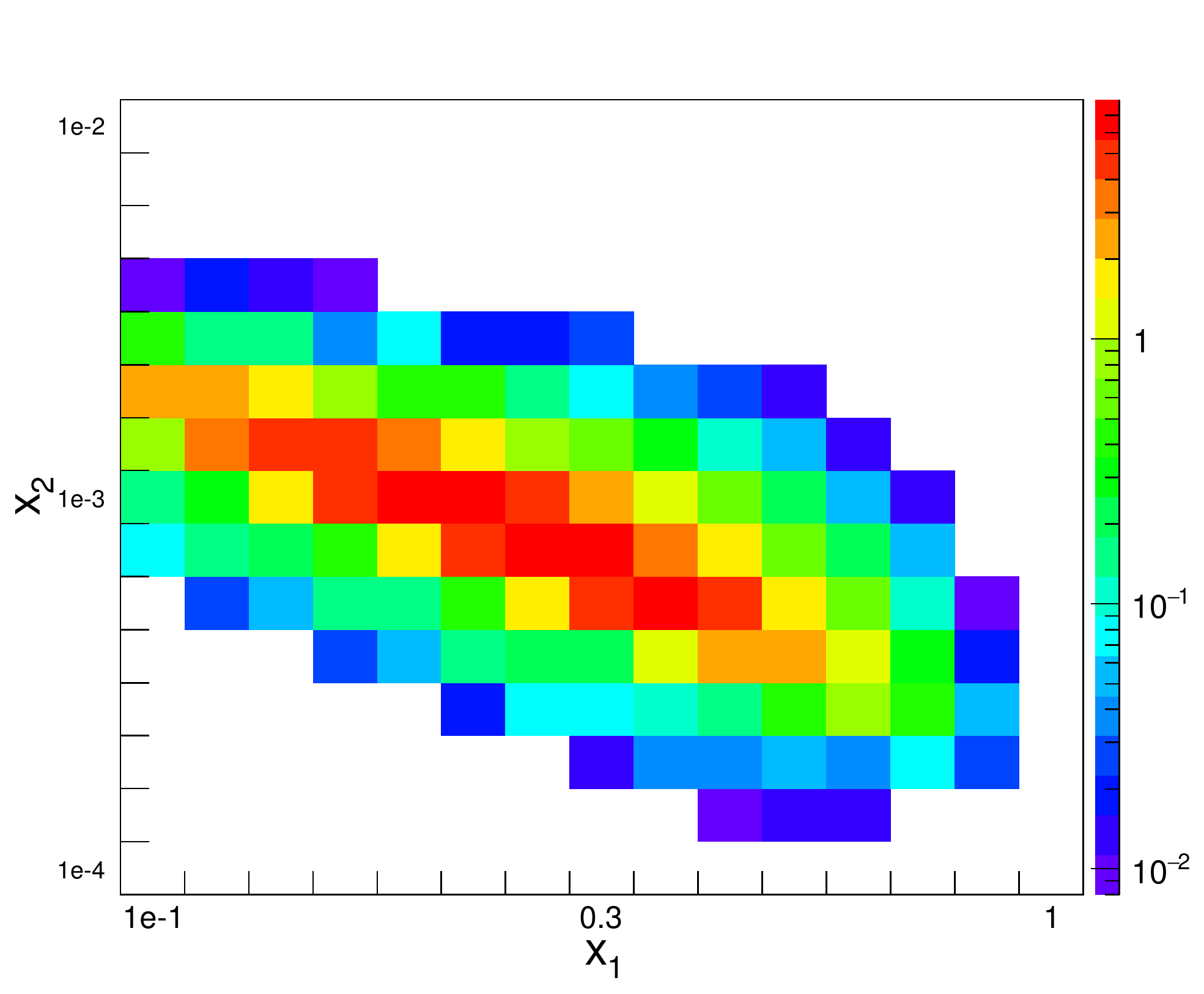}}
\caption{Correlation between longitudinal momentum factions $x$ of the initial state partons
for a selected bin of $Z$ boson transverse momentum: $20<p_T^Z<30$ {\normalfont GeV}.
Distributions for other bins in $p_T^Z$ look similar.}
\label{fig:IS_x1VSx2}
\end{figure*}

In Fig.~\ref{fig:IS_x1VSx2} the correlations between longitudinal momentum fractions of the initial
state partons $x_1$ and $x_2$ are shown for both factorization schemes (HE and hybrid), for the region  $20<p_T^Z<30$ GeV (results for other ranges of $p_T^Z$ are similar).
In the region of forward $Z$ production, as defined by the LHCb measurement,  
one of the partons has a small longitudinal component ($x_2\sim10^{-3}$), whereas the second parton has rather large
values $x_1\sim0.3$.
%
%We also note here that in the hybrid approach, where only one of the initial partons is off-shell
%it is the parton featuring low value of $x$, namely $x_2$.
In the hybrid approach, the initial off-shell parton is the one at low value of $x$, namely $x_2$.

%----------------
%----------------

\begin{figure*}[!htb]
\centering{}
\subfloat[$0<p_T^Z<10\;\mathrm{GeV}$]{
\includegraphics[width=0.33\textwidth]{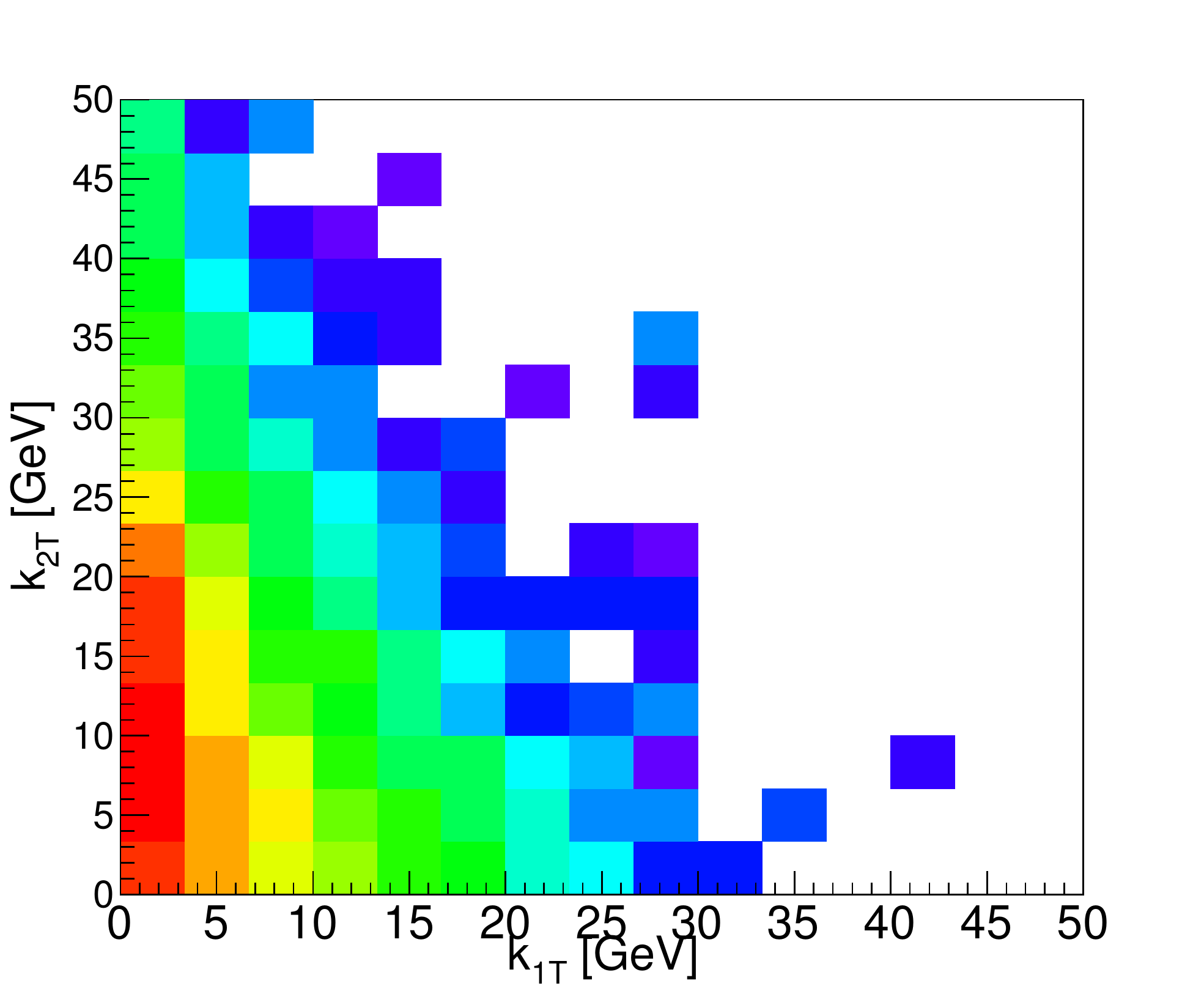}}
\subfloat[$20<p_T^Z<30\;\mathrm{GeV}$]{
\includegraphics[width=0.33\textwidth]{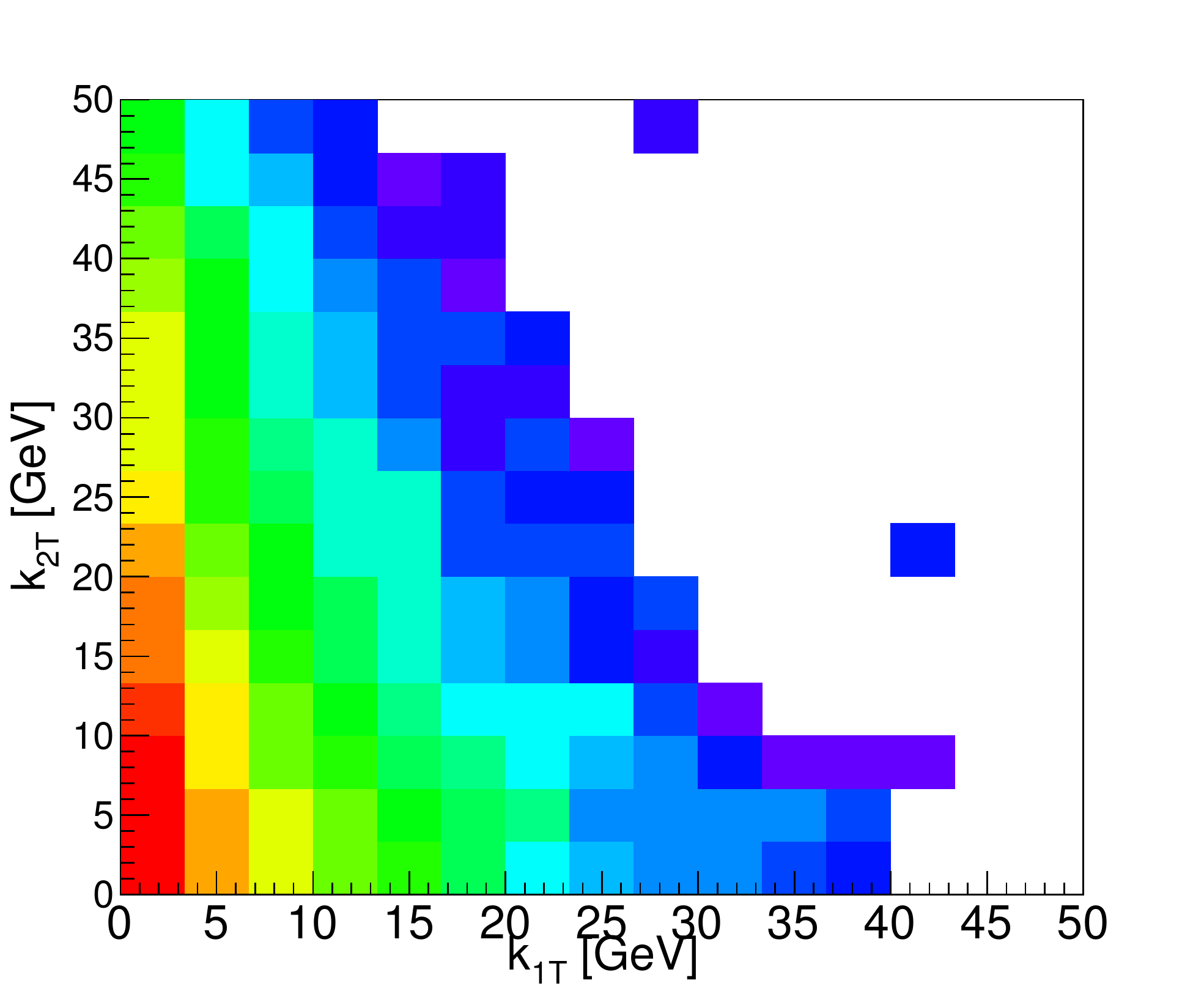}}
\subfloat[$50<p_T^Z<60\;\mathrm{GeV}$]{
\includegraphics[width=0.33\textwidth]{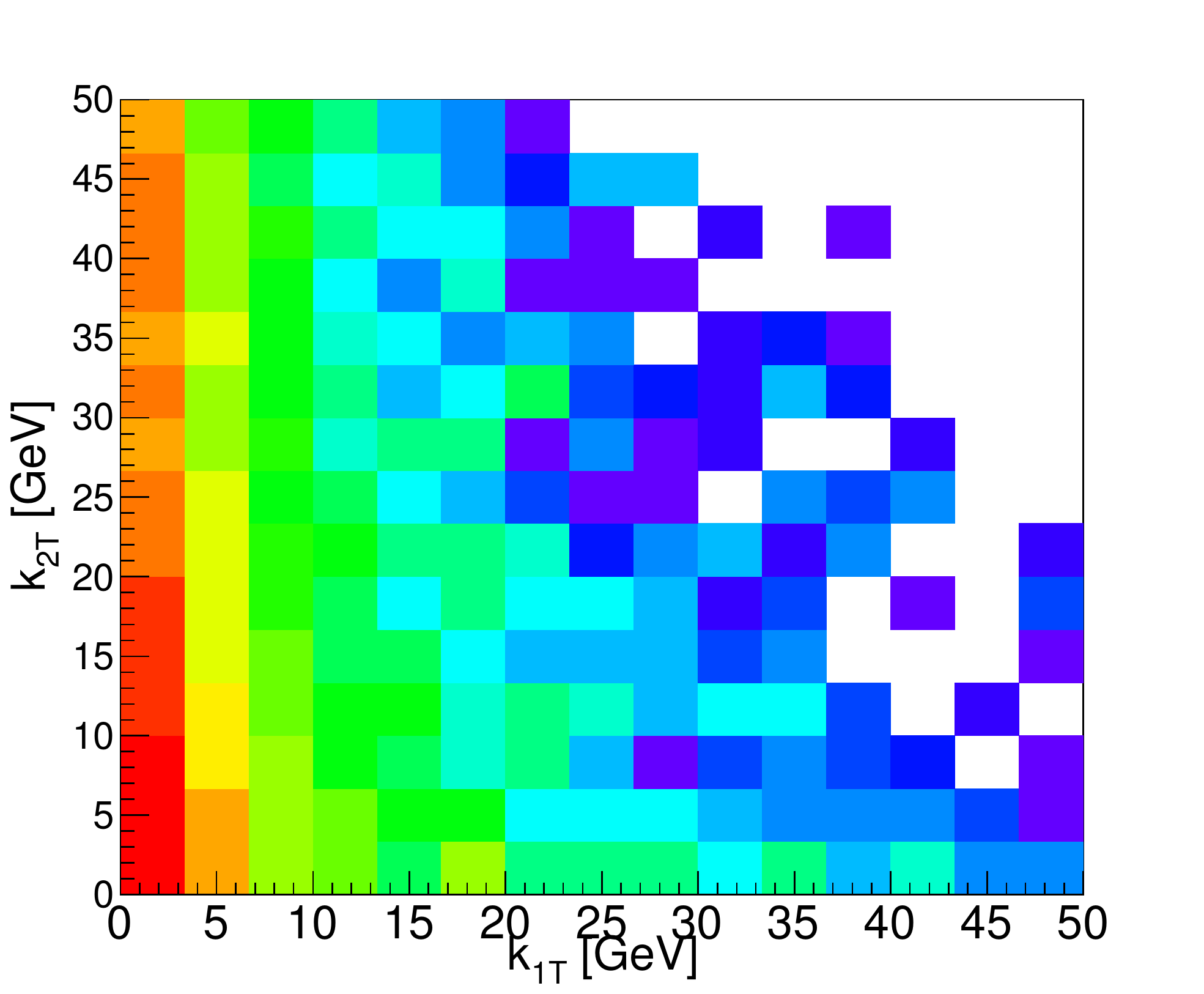}}
\caption{Correlation between transverse momenta of the initial state partons
$k_{1T}$ (with large $x$ value) and $k_{2T}$ (with low $x$ value)
for selected bins of $Z$ boson transverse momentum for the high-energy approach.}
\label{fig:IS_kt1VSkt2}
\end{figure*}

In Fig.~\ref{fig:IS_kt1VSkt2} the correlations between the transverse momenta of the initial state partons  are shown for the case of $k_T$ factorization (in the hybrid approach the $k_{1T}$ momentum is zero)).\footnote{The scale for all the color-map plots is the same as in Fig.~\ref{fig:IS_x1VSx2}.}
One of the transverse momenta, $k_{1T}$ (large $x_1$), is  small ($k_{1T}\lesssim5$ GeV), 
while the other  one, $k_{2T}$ (low $x_2$), has a much broader distribution. The average transverse momentum $k_{2T}$ increases for increasing $p_Z^Z$, as shown in Fig.~\ref{fig:IS_kt1VSkt2}, while the transverse momentum of the other parton stays very small. This observation explains why the predictions in HE (two off-shell partons) 
and hybrid factorization (only one off-shell parton) give  very similar results.
%
%----------------
%----------------
%
%----------------
\begin{figure*}[!htb]
\centering{}
\subfloat[$0<p_T^Z<10\;\mathrm{GeV}$]{
\includegraphics[width=0.33\textwidth]{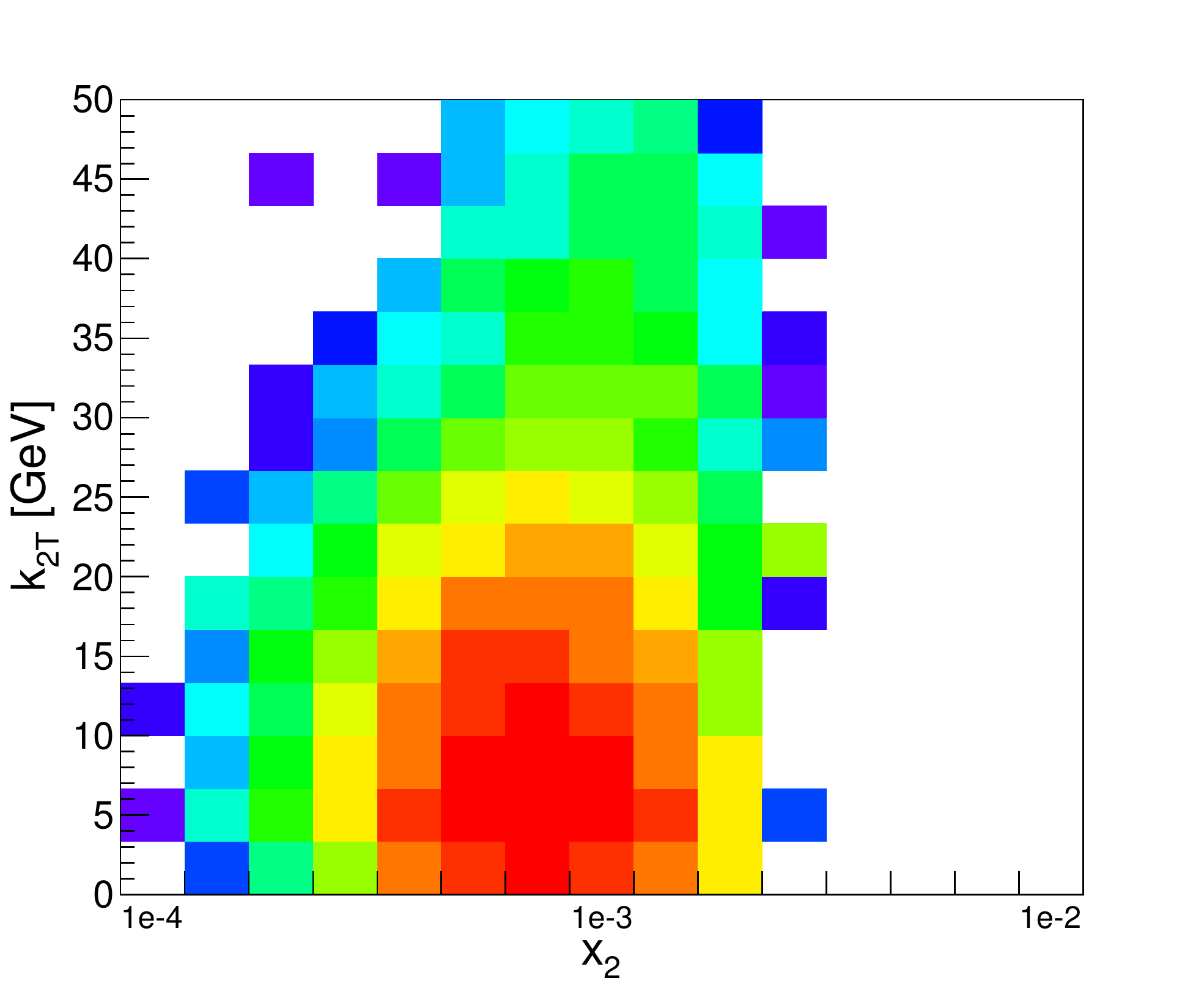}}
\subfloat[$20<p_T^Z<30\;\mathrm{GeV}$]{
\includegraphics[width=0.33\textwidth]{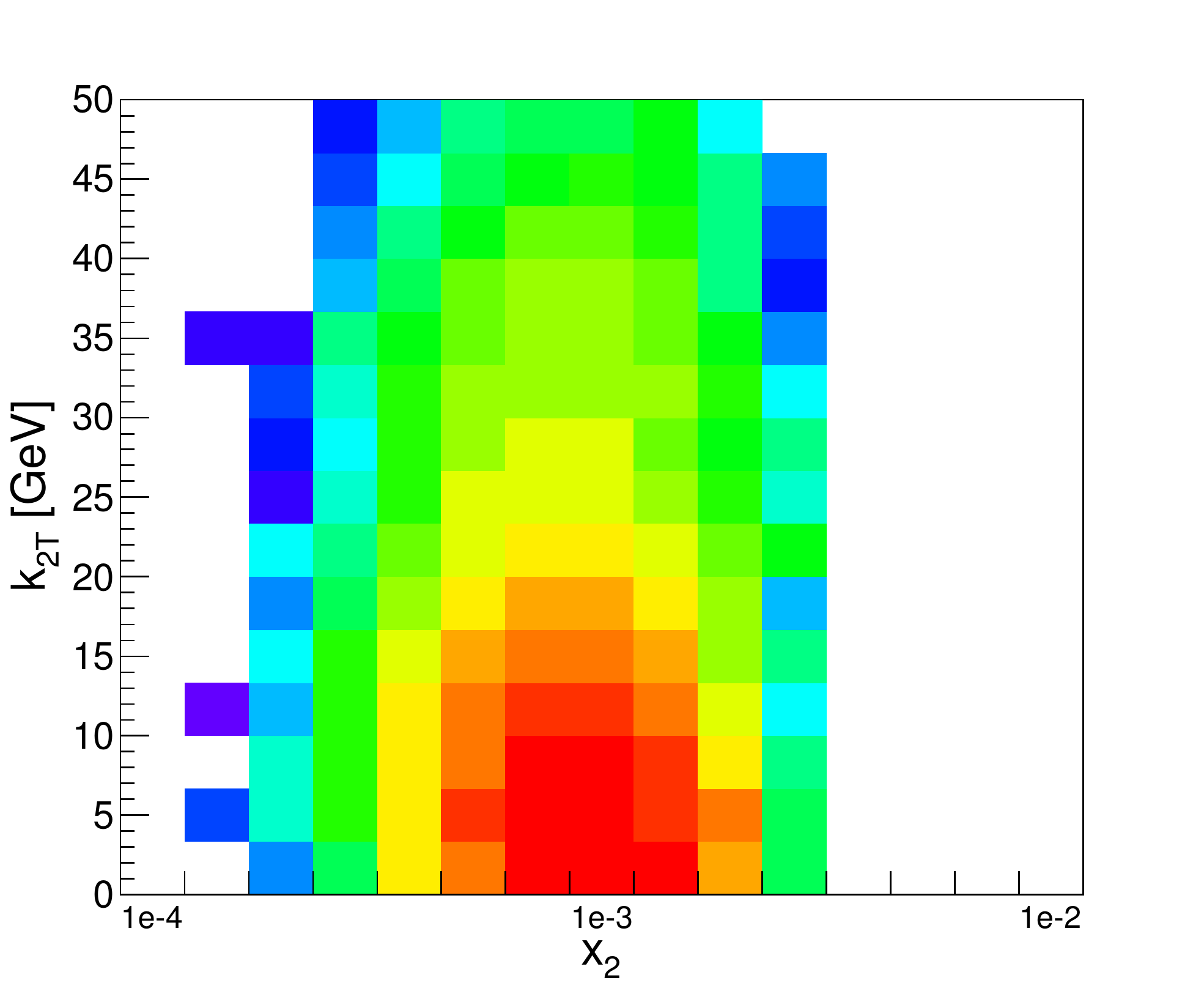}}
\subfloat[$50<p_T^Z<60\;\mathrm{GeV}$]{
\includegraphics[width=0.33\textwidth]{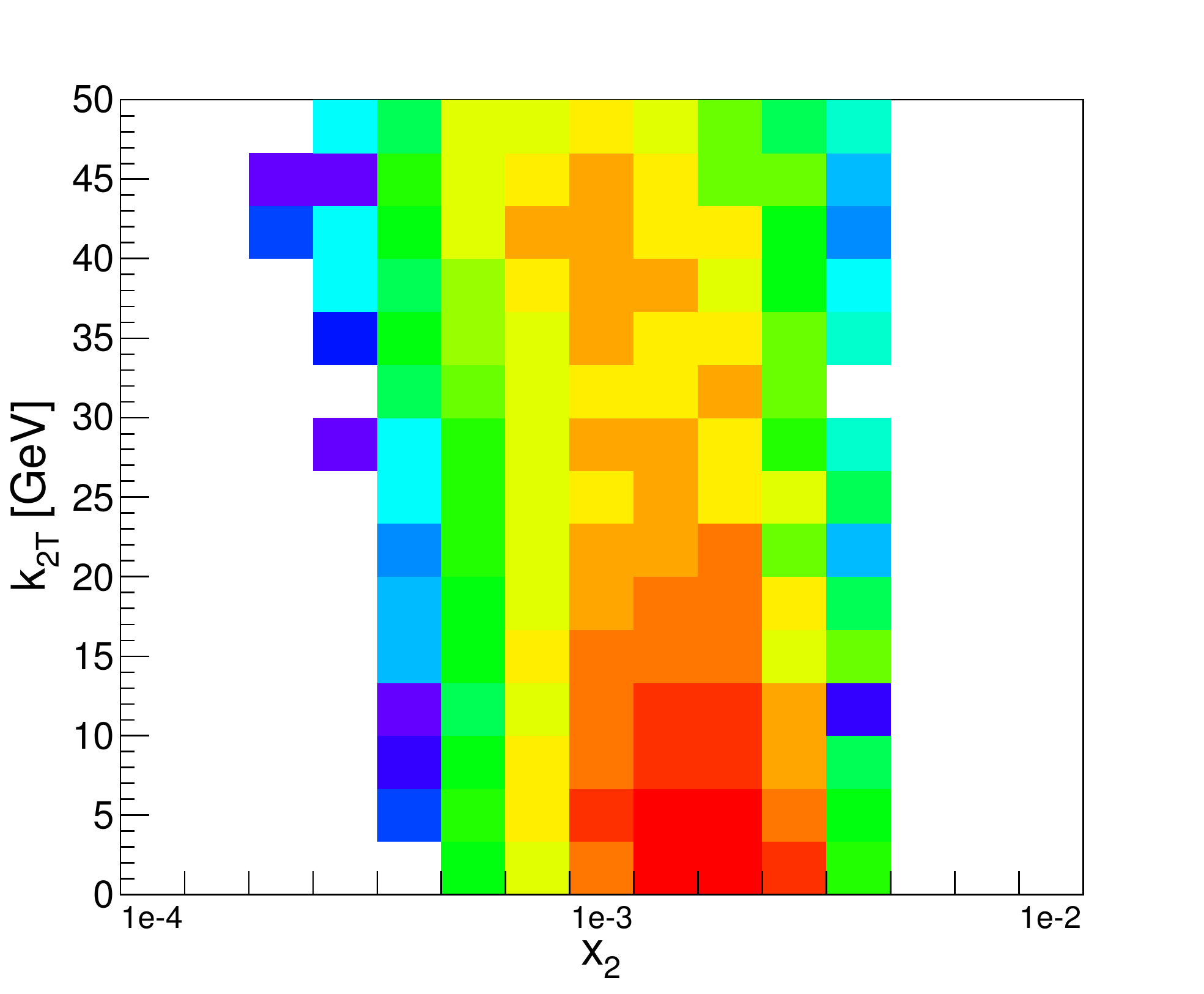}}
\\
\subfloat[$0<p_T^Z<10\;\mathrm{GeV}$]{
\includegraphics[width=0.33\textwidth]{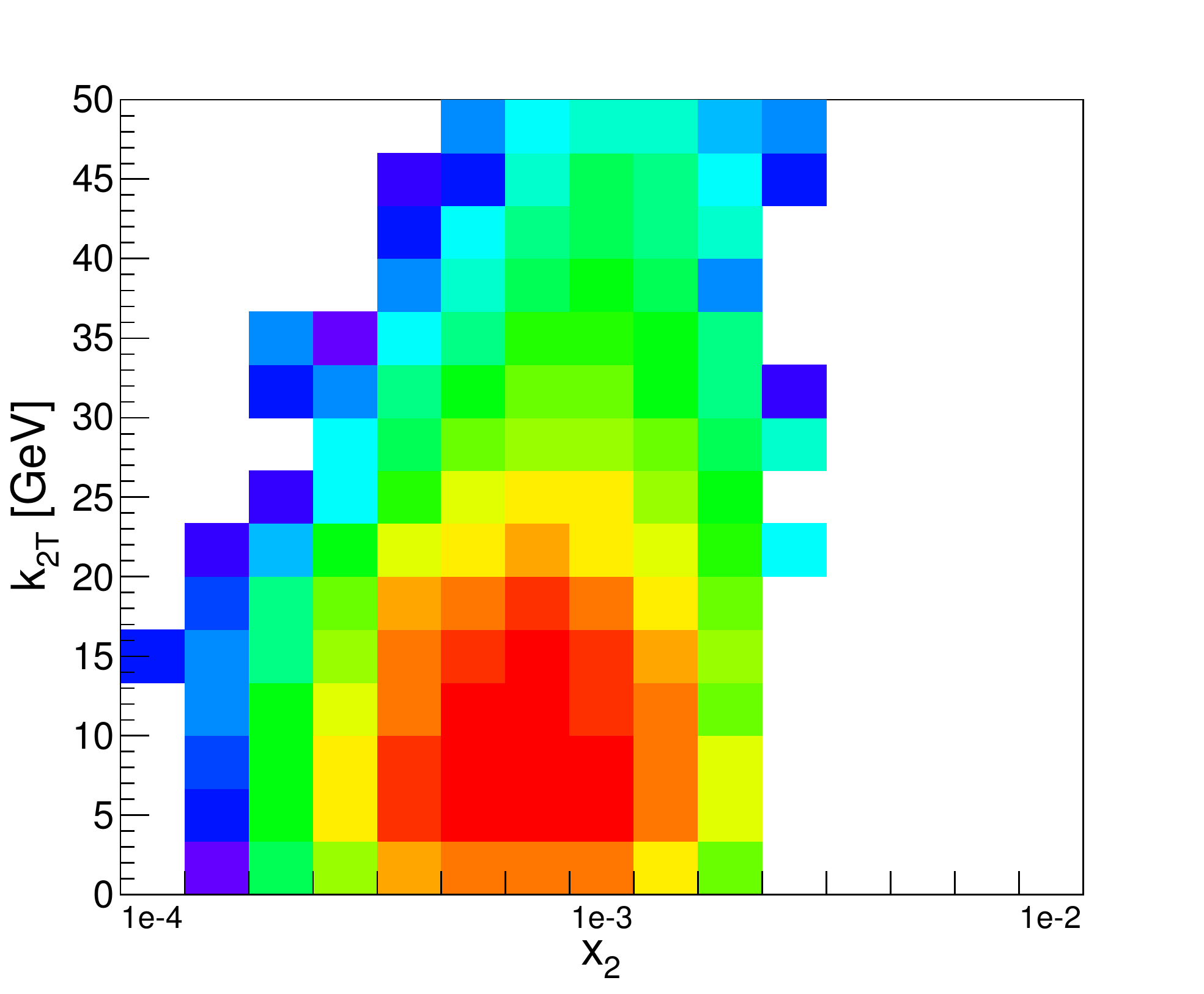}}
\subfloat[$20<p_T^Z<30\;\mathrm{GeV}$]{
\includegraphics[width=0.33\textwidth]{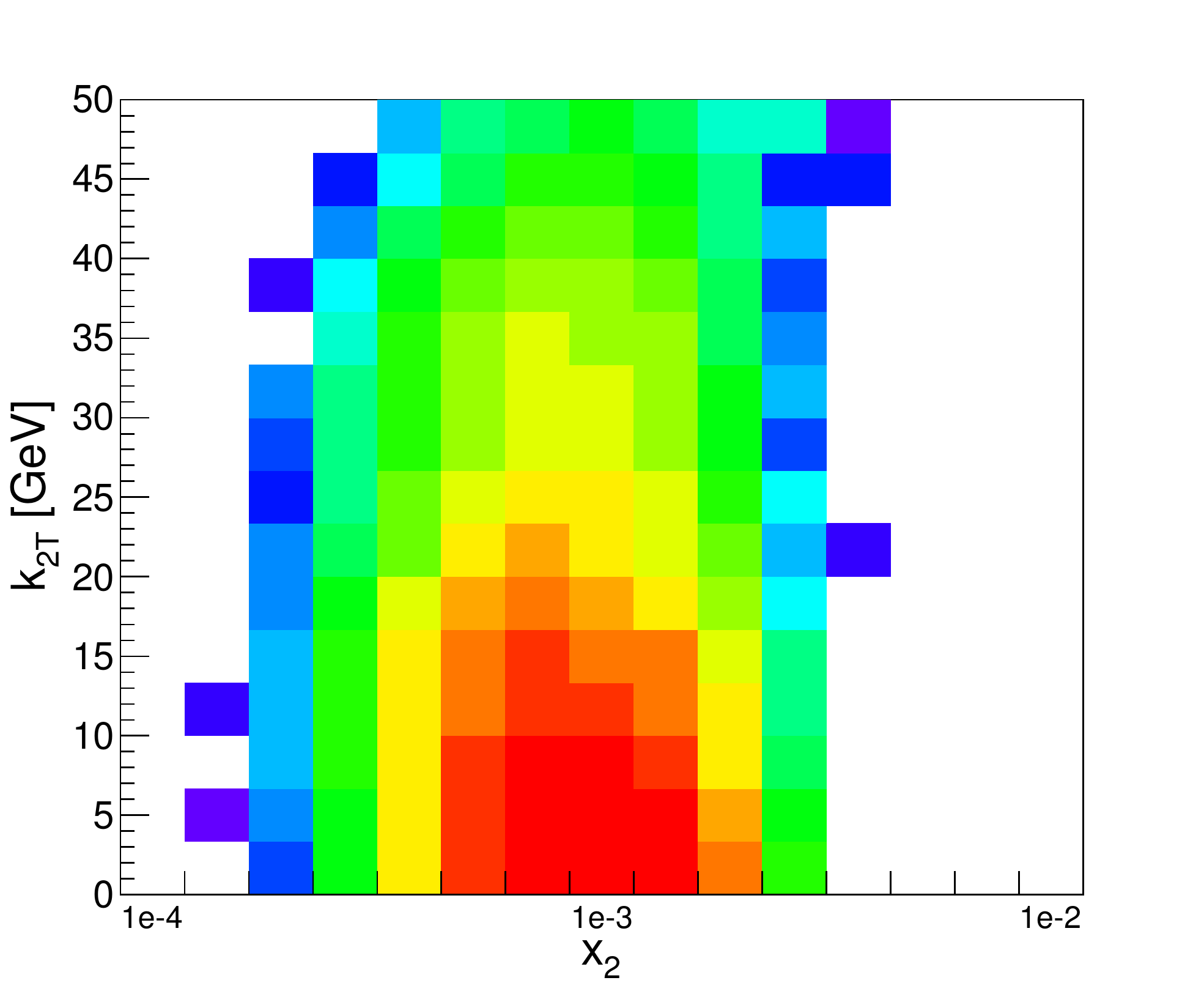}}
\subfloat[$50<p_T^Z<60\;\mathrm{GeV}$]{
\includegraphics[width=0.33\textwidth]{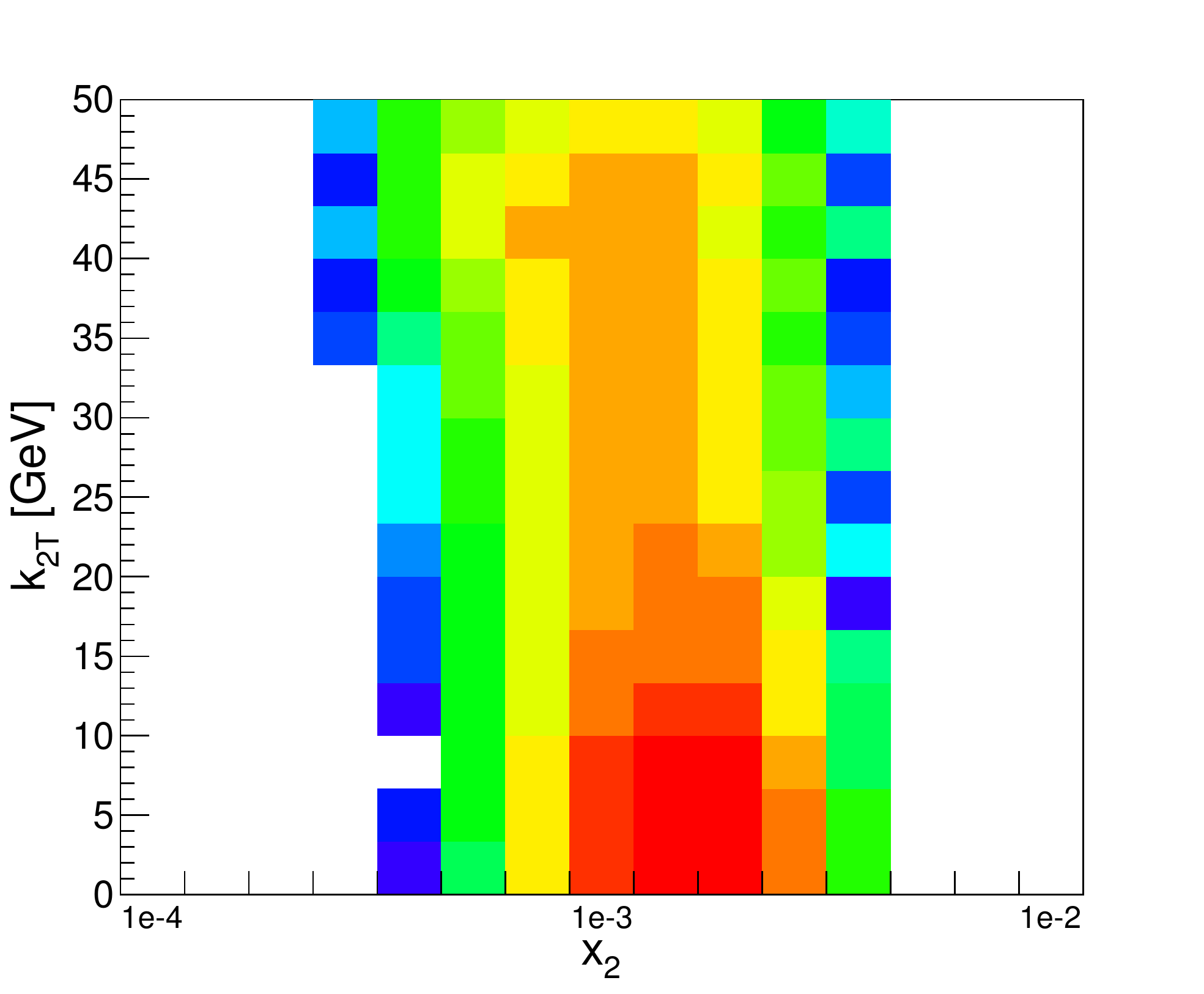}}
\caption{Correlation between longitudinal momentum faction $x_2$ and transverse
momentum of the second initial state parton $k_{2T}$ (low longitudinal momentum fraction)
for selected bins of the $Z$ boson transverse momentum. The upper row displays
results for $k_T$-factorization, lower row for the hybrid approach.}
\label{fig:IS_x2VSkt2}
\end{figure*}
%----------------

In Fig.~\ref{fig:IS_x2VSkt2} the correlation between the longitudinal and the transverse momentum of the low energetic parton is shown, for both $k_T$-factorization and hybrid factorization approaches.
With increasing transverse momentum of the $Z$-boson, the longitudinal momentum $x_2$ also increases. Both factorization approaches show very similar correlations between  $x_2$ and $k_{2T}$ as a function of $p_T^Z$, confirming again the equivalence of $k_T$-factorization and hybrid factorization results in the region of forward $Z$ production.
%

% \clearpage
%%%%%%%%%%%%%%%%%%%%%%%%%%%%%%%%%%%%
\section{Conclusion\label{sec:conclusions}}
%%%%%%%%%%%%%%%%%%%%%%%%%%%%%%%%%%%%
%\comment{What is our main result ?, We should pick up, what we wrote in the abstract.}
We have presented calculations of $Z$+1 jet finals states including transverse momenta of the initial state partons and compared the predictions with measurements of the LHCb experiment. The calculations were performed using the \katie\ parton-level event generator together with initial state parton showers implemented in a new version of \cascade.
We have applied consistently the parton branching transverse momentum dependent parton densities (PB-TMDs) together with two-loop $\alpha_s(m_Z)=0.118$.

The predictions obtained in high-energy and hybrid factorization agree very well with each other
for the forward $Z$ production pointing towards effective equivalence of the two approaches in the
forward region. The predictions obtained in collinear factorisation at leading order for $Z+1$ jet,
supplemented with PB-TMDs and corresponding parton shower show differences to predictions obtained
in high-energy factorization, which come from the different matrix elements used. A comparison of
prediction obtained in high-energy factorisation and a collinear NLO calculation of $Z$+2 jet supplemented with standard parton showers shows very good agreement. The description of the experimental measurement especially at  very small $p_T^Z$ is significantly improved when contributions from multi-parton interactions are included.

The predictions obtained in high-energy factorisation (as well as in hybrid factorization) agree rather well with the measurements of the LHCb collaboration.
Differences are observed in the region of small $p_T^Z$, where the predictions depend significantly on the treatment of multi-parton interactions, which are not yet included in the calculations with high-energy factorization.

We have presented a first consistent comparison of calculations in different factorization approaches, and illustrate the features and advantages of using off-shell matrix elements obtained in $k_T$-factorization.

%%%%%%%%%%%%%%%%%
\section*{Acknowledgments}
%%%%%%%%%%%%%%%%%
Krzysztof Kutak acknowledges the support of Narodowe Centrum Nauki with grant DEC-2017/27/B/ST2/01985.
Michal Deak, Andreas van Hameren and Aleksander Kusina acknowledge the support by FWO-PAS VS.070.16N research grant.
Andreas van Hameren was also partially supported by grant of National Science Center, Poland, No.\ 2015/17/B/ST2/01838.
Mirko Serino is supported by the  Israeli Science Foundation through grant 1635/16, 
by the BSF grants 2012124 and 2014707, by the COST Action CA15213 THOR
and by a Kreitman fellowship by the Ben Gurion University of the Negev. 
Hannes Jung thanks the Polish Science Foundation for the Humboldt Research fellowship during which part of this work was completed.

%%%%%%%%%%%%%%%%
\bibliographystyle{utphys}
\bibliography{references}
%%%%%%%%%%%%%%%%

\end{document}